\documentclass[twocolumn,showpacs,preprintnumbers,amsmath,amssymb]{revtex4}
\usepackage{pifont}

\usepackage{graphicx}
\usepackage{dcolumn}
\usepackage{bm}

\begin{document}

\preprint{APS/123-QED}

\title{Tempered Fractional Feynman-Kac Equation}


\author{Xiaochao Wu$^{1}$}

\author{Weihua Deng$^{1}$}

\author{Eli Barkai$^2$}%

\affiliation{%
  $^1$School of Mathematics and Statistics, Gansu Key Laboratory of Applied Mathematics and Complex
Systems, Lanzhou University, Lanzhou 730000,  P.R. China
\\
$^2$Department of Physics, Advanced Materials and Nanotechnology Institute, Bar Ilan University,  Ramat-Gan 52900,
Israel
 }%



\begin{abstract}
Functionals of Brownian/non-Brownian motions have diverse applications and attracted a lot of interest of scientists. This paper focuses on deriving the forward and backward fractional Feynman-Kac equations describing the distribution of the functionals of the space and time tempered anomalous diffusion, belonging to the continuous time random walk class. Several examples of the functionals are explicitly treated, including the occupation time in half-space, the first passage time, the maximal displacement, the fluctuations of the occupation fraction, and the fluctuations of the time-averaged position.

\end{abstract}

\maketitle

\section{Introduction}
Normal diffusion describes the Brownian dynamics characterized by a large number of small events, e.g., the motion of pollen grains in water. However, in many cases, the (rare) large fluctuations result in the non-Brownian motion, anomalous diffusion, being carefully studied in physics \cite{Metzler:00, Metzler:01}, finance \cite{Scalas:02}, hydrology \cite{Schumer:03}, and many other fields. In particular, based on the continuous time random walk (CTRW) model, the corresponding fractional Fokker-Planck or diffusion equations are derived (see the review article \cite{Metzler:00} and numerical methods \cite{Deng:08}).

Tempered anomalous diffusion describes the very slow transition from anomalous to normal diffusion and it has many applications in physical, biological, and chemical processes \cite{Bruno:00,Meerschaert:11,Stanislavsky:08,Stanislavsky:14}; and for numerical methods see \cite{Baeumer:10, Gajda:10}. In some cases, the transition even does not appear at all in the observation time because of the finite lifespan of the particles or the finite observation time of the experimentalist. As a generalization of the Brownian walk, the CTRW model allows the incorporation of the waiting time distribution $\psi(t)$ and the general jump length distribution $\eta(x)$. The CTRW model describes the normal diffusion if $\psi(t)$ has bounded first moment and $\eta(x)$ bounded second moment, e.g., $\psi(t)$ is exponential distribution and $\eta(x)$ is Gaussian distribution. Anomalous diffusion is characterized by the CTRW model with the waiting time distribution $\psi(t)$ having divergent first moment and/or the jump length distribution $\eta(x)$ having divergent second moment, e.g.,  $\psi(t)\simeq t^{-\alpha-1}$ $( 0<\alpha<1)$ and/or $\eta(x)\simeq  |x|^{-\beta-1}$ $ ( 0<\beta<2)$. Sometimes the more reasonable/physical choice for $\psi(t)$ is to make it have finite first moment; similarly, sometimes the bounded physical space implies that $\eta(x)$ should have finite second moment. These can be realized by truncating the heavy tail of the power-law distribution \cite{Sokolov:01}. The tempered anomalous diffusion is described by the CTRW model with truncated power-law waiting time and/or jump length distribution(s). In this paper, we use the exponentially truncated stable distribution (ETSD) \cite{Koponen:11,Nakao:11} waiting time $\psi(t,\lambda)$ and exponentially truncated jump length $\eta(x)$. The exponential tempering offers technical advantages since the tempered process is still an infinitely divisible L\'{e}vy process  \cite{Meerschaert:00}.

%
%
%
%
%

It is well known that there are many physical quantities used to describe the motion features of a Brownian particle. An example one, Brownian functionals, being defined as $A=\int_{0}^{t}U[x(\tau)]d\tau$, where $U(x)$ is a prescribed function and $x(t)$ is a trajectory of a Brownian particle. Here $A$ is a random variable since $x(t)$ is a stochastic process. Functionals of diffusion motion have diverse applications and have been well studied, including functionals of Brownian motion \cite{Kac:01} and non-Brownian motion \cite{Turgeman:00, Carmi:00, Carmi:01}. In particular, based on the sub-diffusive CTRW, a widely investigated process being continually used to characterize the motion of particles in disordered systems \cite{Havlin:00, Montroll:00, Scher:00}, the fractional Feynman-Kac equation is derived \cite{Turgeman:00}. Taking the tempered power-law function $\psi(t,\lambda)$ as the waiting time distribution in the CTRW model, in this paper we derive the forward and backward Feynman-Kac equations governing the distribution of the functionals of the tempered anomalous diffusion; and the tempered fractional substantial derivative \cite{Friedrich:00} is used in the equations. The derivations include several cases: random walk on lattice; random walk on lattice with forces; power-law jump distribution; tempered power-law jump distribution. After deriving the equations, several concrete examples of the functionals of the tempered anomalous diffusion are analytically and explicitly analyzed, covering the occupation time in half-space \cite{Barkai:04}, the first passage time, the maximal displacement, the fluctuations of the occupation fraction, the fluctuations of the time-averaged position, and the ergodic behavior of the particle.

The paper is organized as follows. In Section \uppercase \expandafter {\romannumeral 2}, we derive the forward and backward fractional Feynman-Kac equations for the functionals of the tempered anomalous diffusion with different jump length distributions. In Section \uppercase \expandafter {\romannumeral 3}, we present the solutions of the equations for a few functionals of interest of free particles. Then we discuss the occupation time in half-space, the first passage time, the maximal displacement, and the fluctuations of the occupation fraction with the obtained solutions. In Section \uppercase \expandafter {\romannumeral 4}, we investigate the fluctuations of the time-averaged position. The paper is concluded with some comments in the last section.



\section{Derivation of the Equations} \label{Sec2}
\subsection{Model}
\begin{figure}
  \centering
  \includegraphics[height=4cm,width=9cm]{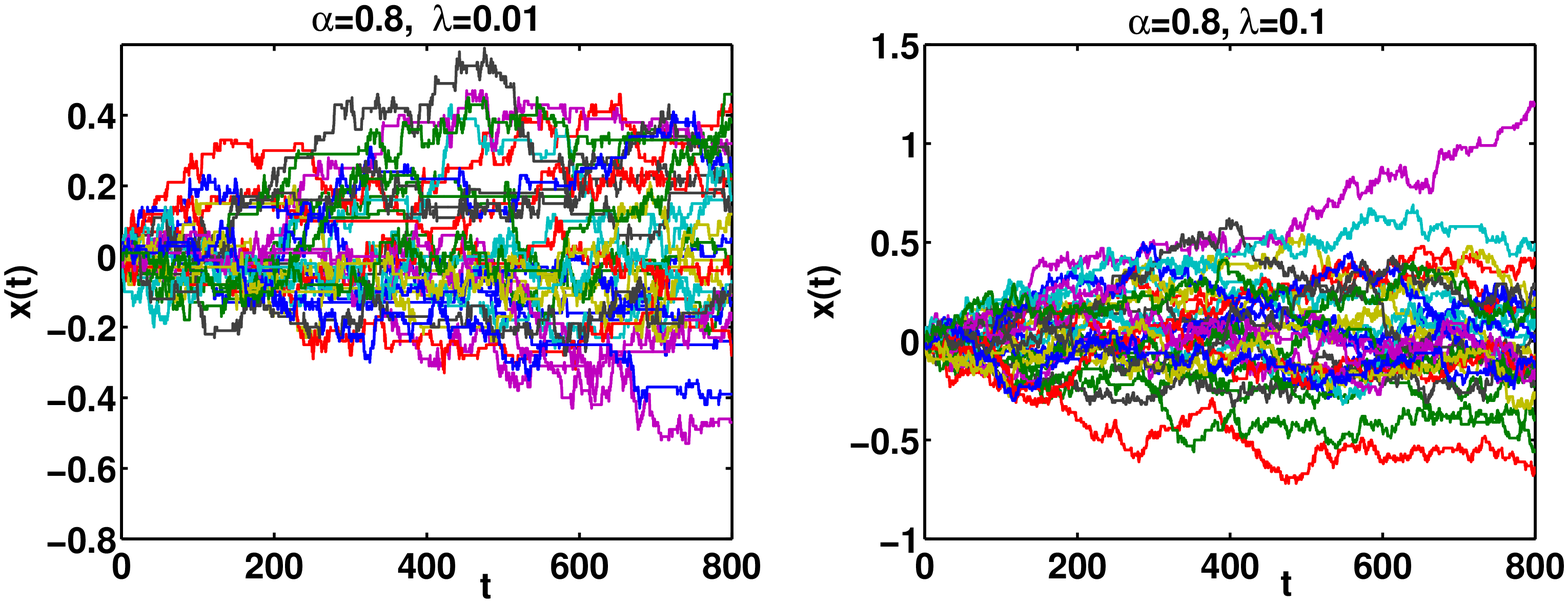}\\
  \caption{Trajectories of $30$ particles, moving on the lattice with the tempered waiting times, where $a=0.01$,  $K_{\alpha}=1/2$, and $\lambda=0.01$ and $0.1$.}\label{fig.9}
\end{figure}

We use the CTRW model with tempered power-law waiting time distribution as the underlying process leading to tempered anomalous diffusion, which characterizes the slow transition from the anomalous to normal diffusion which is controlled by the parameter $\lambda$ (see Fig. \ref{fig.9}).
 First, we consider the CTRW on a lattice, i.e., a particle is placed on an infinite one-dimensional lattice with spacing $ a $ and is allowed to jump to its nearest neighbors only. The probabilities of jumping left $ L(x) $ and right $
R(x) $ depend on $ F(x) $, the force at $ x $ (see subsection {\bf C} for a derivation of these probabilities). If $F(x)=0$, then $ R(x)=L(x)=1/2 $. Waiting times between jump events are independent identically distributed (i.i.d.)
random variables with exponentially truncated stable distribution (ETSD) $ \psi(t,\lambda)$, and are independent of the external force.
This ETSD is useful for rigorous analysis of diffusion behavior because it is an infinitely divisible distribution, and thus its distribution or characteristic function can be explicitly derived. The Laplace transform for $\psi(t,\lambda)$ is given by \cite{Feller:11}
$$ e^{\hat{\phi}(s,\lambda)}=\int_{0}^{+\infty}\psi(t,\lambda)e^{-st}dt, $$
where $\hat{\phi}(s,\lambda)=-B_{\alpha}(\lambda+s)^{\alpha}+B_{\alpha} \lambda^{\alpha}$. Hence, the Laplace transform (for small $s$ and $\lambda$) of ETSD $\psi(t,\lambda)$ results in
\begin{equation} \label{LaptemperedWTPDF}
\begin{split}
\hat{\psi}(s,\lambda)& =e^{-B_{\alpha}(\lambda+s)^{\alpha}+B_{\alpha} \lambda^{\alpha}} \\
& \simeq 1-B_{\alpha}(\lambda+s)^{\alpha}+B_{\alpha} \lambda^{\alpha}.
\end{split}
\end{equation}

The process starts at $ x=x_{0} $, and the particle waits at $ x_{0} $ for time $ t $ drawn from $ \psi(t,\lambda) $ and then jumps to either $ x_{0}+a $ (with probability $R(x_{0})$) or $ x_{0}-a $ (with probability $L(x_{0})$), after which the process is renewed. Furthermore, we consider the more general cases, i.e., instead of being a fixed number, the stepsizes are random variables, subject to power-law or tempered power-law distribution.


\subsection{A random walk on a one-dimensional lattice}
We now consider the CTRW on a lattice. Let $ G(x,A,t)$ be the joint probability density function (PDF) of finding the particle at position $x$ and time $t$ with the functional value $A$. Here the functional $A= \int_0 ^t U[x(t)] {\rm d} t $ as usual.  And $ G(k,p,s)$ is the Fourier transformation $x\rightarrow k$,  $ A\rightarrow p $, and Laplace transformation $ t\rightarrow s $ of $ G(x,A,t) $. In this subsection, based on the CTRW model with ETSD describing the tempered anomalous diffusion, we derive the forward and backward tempered fractional Feynman-Kac equations governing $G(x,p,t)$.
\\
{\em Derivation of the forward tempered fractional Feynman-Kac equation from the random walk on a lattice.} --- For a random walk on a lattice with a general given waiting time PDF $\psi(t,\lambda)$, the following formal solution was obtained by Carmi et al (see Appendix  A) \cite{Turgeman:00}
\begin{equation} \label{GkpsLatt}
\begin{array}{c}
G(k,p,s)=\frac{1-\hat{\psi}[s+pU(-i\frac{\partial}{\partial k}),\lambda]}{s+pU(-i\frac{\partial}{\partial k})}\cdot \frac{1}{1-\cos(ka) \hat{\psi}[s+pU(-i\frac{\partial}{\partial k}),\lambda]},
\end{array}
\end{equation}
where $ U(x) $ is a prescribed function. Recently, Cairoli and Baule \cite{Baule:00} derived a generalized Feynman-Kac equation for CTRW functionals; one of our goals is to now show that this equation can be directly derived from Eq. (\ref{GkpsLatt}), thus hopefully clarifying better the meaning of the latter equation given in \cite{Turgeman:00}. Note that in Eq. (\ref{GkpsLatt}), $\cos(ka)$ is the Fourier transform of the jump length distribution, which on a lattice is a sum of two delta functions of step sizes $\pm a$.

Substituting Eq. (\ref{LaptemperedWTPDF}) into Eq. (\ref{GkpsLatt}) and using $ \cos(ka)\simeq 1-\frac{a^{2}k^{2}}{2}$ for the long wavelength $ k\rightarrow 0$ corresponds to large $x$ (or the continuous limit $a\rightarrow0$) as is well known, then
%
\begin{widetext} \begin{equation} \label{eq4}
G(k,p,s)\simeq\frac{B_{\alpha}[\lambda+s+pU(-i\frac{\partial}{\partial k})]^{\alpha}-B_{\alpha} \lambda^{\alpha}}{s+pU(-i\frac{\partial}{\partial k})}\times \frac{1}{\frac{a^{2}k^{2}}{2}+B_{\alpha}[\lambda+s+pU(-i\frac{\partial}{\partial k})]^{\alpha}-B_{\alpha} \lambda^{\alpha}}.
\end{equation}
After some rearrangements to Eq. (\ref{eq4}), we have
\begin{equation}\label{G1kps}
\begin{split}
\left\{\frac{a^{2}k^{2}}{2B_{\alpha}}\left[\lambda+s+pU\left(-i\frac{\partial}{\partial k}\right)\right]^{1-\alpha}+\left[\lambda+s+pU\left(-i\frac{\partial}{\partial k}\right)\right]-\lambda^{\alpha}\left[\lambda+s+pU\left(-i\frac{\partial}{\partial k}\right)\right]^{1-\alpha}\right\}G(k,p,s)-1
\\
=\frac{\lambda-\lambda^{\alpha}\left[\lambda+s+pU\left(-i\frac{\partial}{\partial k}\right)\right]^{1-\alpha}}{s+pU(-i\frac{\partial}{\partial k})}.
\end {split}
\end{equation}
Inverting to the space-time domain $ k\rightarrow x $ and $ s\rightarrow t $ , from the well-known Fourier transformation $ \mathcal{F} \{ xf(x);k \}=-i\frac{\partial}{\partial k}\hat{f}(k)$; the operator $U(-i\frac{\partial}{\partial k})$ on the right hand side (rhs) of Eq. (\ref{G1kps}) is operating on $1$, which means that we use the Taylor expansion $U(-i\frac{\partial}{\partial k})=\displaystyle \sum_{n= 0}^{\infty}c_{n}\left(-i\frac{\partial}{\partial k}\right)^{n}$ and thus we consider the analytical functions
$U(x)$. Note that the order of the terms is important: for instance, $k^{2}$ does not commute with $U(-i\frac{\partial}{\partial k})$; thus we finally find the tempered fractional Feynman-Kac equation:
\begin{equation} \label{1stFTEq}
\frac{\partial}{\partial t}G(x,p,t)=
\left[\lambda^{\alpha}D_{t}^{1-\alpha,\lambda}-\lambda\right]\left[G(x,p,t)-e^{-pU(x)t}\delta(x)\right]-pU(x)G(x,p,t)
+K_{\alpha}\frac{\partial^{2}}{\partial x^{2}}D_{t}^{1-\alpha,\lambda}G(x,p,t),
\end{equation}
with the initial condition $ G(x,A,t=0)=\delta(x)\delta(A) $ or $ G(x,p,t=0)=\delta(x)$, where $ \delta(\cdot) $ is the Dirac delta function and
\begin{equation}\label{Kalpha}
K_{\alpha}=\frac{a^{2}}{2B_{\alpha}},
\end{equation}
with units $m^{2}/\sec^{\alpha}$, is finite for $a\rightarrow0, B_{\alpha}\rightarrow 0$. Eq. (\ref{Kalpha}) is a generalized Einstein relation for tempered motion. 
In Laplace space, $ D_{t}^{1-\alpha,\lambda}\rightarrow [\lambda+s+pU(x)]^{1-\alpha}$; and in $t$ space, the tempered fractional substantial derivative
\begin{equation} \label{temperedFSD}
D_{t}^{1-\alpha,\lambda} G(x,p,t)=\frac{1}{\Gamma(\alpha)}
\left[\lambda+pU(x)+\frac{\partial}{\partial t}\right] \int_0^t
\frac{e^{-(t-\tau)\cdot (\lambda+pU(x))}}{(t-\tau)^{1-\alpha}}
G(x,p,\tau) d \tau.
\end{equation}
\end{widetext}
Thus, due to the long waiting times, the evolution of $ G(x,p,t) $ is non-Markovian and depends on the entire history. Throughout the paper, the operator $D_{t}^{1-\alpha,\lambda}$ is defined as in Eq. (\ref{temperedFSD}).
\\
{\em (a).} With the case that $ \lambda $ is finite but $ \alpha=1 $, the equation leads to
$$ \frac{\partial}{\partial t}G(x,p,t)=K_{1}\frac{\partial^{2}}{\partial x^{2}}G(x,p,t)-pU(x)G(x,p,t). $$
This is simply the famed Feynman-Kac equation \cite{Kac:01}. Namely,  exponentially truncation has no effects on normal diffusion. As well known the Feynman-Kac equation is the imaginary time Schr\"{o}dinger equation, where $U(x)$ serves as the potential field.
\\
{\em (b).} When $ \lambda=0$ , Eq. (\ref{1stFTEq}) reduces to the imaginary time fractional Schr\"{o}dinger equation \cite{Turgeman:00}, namely the traditional Feynman-Kac equation.
\begin{equation} \label{eq7}
\begin{array}{c}
\frac{\partial}{\partial t}G(x,p,t)=K_{\alpha}\frac{\partial^{2}}{\partial x^{2}}D_{t}^{1-\alpha}G(x,p,t)-pU(x)G(x,p,t).
\end{array}
\end{equation}
{\em (c).} Furthermore, if $p=0$, Eq. (\ref{eq7}) turns to the fractional diffusion equation \cite{Metzler:00}:
$$\frac{\partial}{\partial t}G(x,t)=K_{\alpha}\frac{\partial^{2}}{\partial x^{2}}D_{RL,t}^{1-\alpha}G(x,t),$$
where $D_{RL,t}^{1-\alpha}$ is the Riemann-Liouville fractional derivative operator.
\\
{\em (d).} When $p=0$ and $\lambda$ is finite, Eq. (\ref{1stFTEq}) becomes:
\begin{equation}
\begin{split}
\frac{\partial}{\partial t}G(x,t)&=
\left[\lambda^{\alpha}D_{t}^{1-\alpha,\lambda}-\lambda\right]\left[G(x,t)-\delta(x)\right] \\
&+K_{\alpha}\frac{\partial^{2}}{\partial x^{2}}D_{t}^{1-\alpha,\lambda}G(x,t),
\end{split}
\end{equation}
where $$D_{t}^{1-\alpha,\lambda} G(x,t)=\frac{1}{\Gamma(\alpha)}
\left[\lambda+\frac{\partial}{\partial t}\right] \int_0^t
\frac{e^{-(t-\tau)\lambda}}{(t-\tau)^{1-\alpha}}
G(x,\tau) d \tau.$$
This is a diffusion equation for tempered CTRW processes.
\begin{widetext}
{\em Another formulation for Eq. (\ref{1stFTEq}).} --- Rearranging Eq. (\ref{eq4}) leads to
\begin{equation}
\left[s+pU\left(-i\frac{\partial}{\partial k}\right)\right]G(k,p,s)-1
=-\frac{a^{2}k^{2}}{2B_{\alpha}}\left[s+pU\left(-i\frac{\partial}{\partial k}\right)\right]\frac{G(k,p,s)}{\hat{\Phi}[s+pU(-i\frac{\partial}{\partial k}),\lambda]},
\end{equation}
where $\hat{\Phi}(s,\lambda)=(\lambda +s)^{\alpha}-\lambda^{\alpha}$. Inverting to the time-space domain $ s\rightarrow t $ and $ k\rightarrow x $, from the well-known Fourier transformation $ \mathcal{F} \{ g_{1}(x)g_{2}(x);k \}=\hat{g_{1}}(-i\frac{\partial}{\partial k})\hat{g_{2}}(k)$, the tempered fractional Feynman-Kac equation is obtained as
\begin{equation} \label{a1stFTEq}
 \frac{\partial}{\partial t}G(x,p,t)+pU(x)G(x,p,t)=  \\ K_{\alpha}\frac{\partial^{2}}{\partial x^{2}}\left[\frac{\partial}{\partial t}+pU(x)\right]\int_{0}^{t}K(t-\tau,\lambda)e^{-pU(x)(t-\tau)}G(x,p,\tau)d\tau,
\end{equation}
\end{widetext}
with the initial condition $ G(x,A,t=0)=\delta(x)\delta(A) $ or $ G(x,p,t=0)=\delta(x)$, where the memory kernel is related to $\Phi$ by $ \hat{K}(s,\lambda)=\hat{\Phi}(s,\lambda)^{-1} $ and given by $K(t,\lambda)=e^{-\lambda t}t^{\alpha-1}E_{\alpha,\alpha}((\lambda t)^{\alpha})$. Here $E_{\alpha,\alpha}(\cdot)$ is the Mittag-Leffler function. It can be noted that Eq. (\ref{a1stFTEq}) is exactly the same as Eq. (15) of \cite{Baule:00}.

{\em Derivation of the backward tempered Feynman-Kac equation from the random walk on a lattice.} --- Now we derive a backward equation which turns out to be very useful. In some cases we may be just interested in the distribution of $ A $, so integrating $ G(x,A,t) $ over all $ x $ is necessary. Therefore, it would be convenient to obtain an equation for $G_{x_{0}}(A,t)$, which is the PDF of the functional $A$ at time $t$ for a process starting at $x_{0}$.
According to the CTRW model, the particle starts at $ x=x_{0} $; after its first jump at time $\tau$, it is at either $x_{0}+a$ or $x_{0}-a$. Alternatively, the particle doesn't move at all during the measurement time $(0,t)$. Translating this process to an equation, there exists \cite{Turgeman:00}
\begin{widetext}
\begin{equation} \label{eq9}
G_{x_{0}}(A,t)=\int_{0}^{t}d\tau \psi(\tau,\lambda)\frac{1}{2}\left\{G_{x_{0}+a}[A-\tau U(x_{0}),t-\tau]+G_{x_{0}-a}[A-\tau U(x_{0}),t-\tau]\right\}+W(t,\lambda)\delta[A-tU(x_{0})],
\end{equation}
\end{widetext}
where $\tau U(x_{0})$ is the contribution to $A$ from the pausing time on $x_{0}$ in the time interval $(0,\tau)$; the last term on the right hand side of Eq. (\ref{eq9}) shows motionless particles, for which $ A(t)=tU(x_{0}) $; and  $W(t,\lambda)=1-\int_{0}^{t}\psi(\tau,\lambda)d\tau$ is the probability that particle remained motionless on its initial location. The Laplace transform of $W(t,\lambda)$ follows from the form for the Laplace transform of an integral \cite{Arfken:00} and reads $\hat{W}(s,\lambda)=\frac{1-\hat{\psi}(s,\lambda)}{s}$, here $ \hat{\psi}(s,\lambda) $ is also given by Eq. (\ref{LaptemperedWTPDF}). Taking the Laplace transform $t\rightarrow s$ and Fourier transforms $x_{0}\rightarrow k$ and $A\rightarrow p$, we have
\begin{widetext}
\begin{equation}\label{beq}
\begin{split}
  G_{k}(p,s) & =\hat{\psi}\left[pU\left(-i\frac{\partial}{\partial k}\right)+s,\lambda\right]\cos(ka)G_{k}(p,s)+\hat{W}\left[pU\left(-i\frac{\partial}{\partial k}\right)+s,\lambda \right]\delta(k) \\
             & =\left\{1-B_{\alpha}\left[\lambda+pU\left(-i\frac{\partial}{\partial k}\right)+s \right]^{\alpha}+B_{\alpha}\lambda^{\alpha}\right\}\cos(ka)G_{k}(p,s)+\frac{B_{\alpha}[\lambda+pU\left(-i\frac{\partial}{\partial k}\right)+s]^{\alpha}-B_{\alpha}\lambda^{\alpha}}{pU\left(-i\frac{\partial}{\partial k}\right)+s}\delta(k).
\end{split}
\end{equation}
Rearranging the expressions and taking approximation $ k\rightarrow 0 $, $ \cos(ka)\simeq 1-\frac{a^{2}k^{2}}{2} $ in the last equation we find
\begin{equation} \label{eq11}
\begin{split}
\left[\lambda+pU\left(-i\frac{\partial}{\partial k}\right)+s\right]^{1-\alpha}\frac{a^{2}k^{2}}{2B_{\alpha}}G_{k}(p,s)  +  \left\{\left[\lambda+pU\left(-i\frac{\partial}{\partial k}\right)+s\right]-\lambda^{\alpha}\left[\lambda+pU\left(-i\frac{\partial}{\partial k}\right)+s\right]^{1-\alpha}\right\}G_{k}(p,s) \\
  -\delta(k) =\frac{\lambda-\lambda^{\alpha}[\lambda+pU(-i\frac{\partial}{\partial k})+s]^{1-\alpha}}{pU(-i\frac{\partial}{\partial k})+s}\delta(k).
\end{split}
\end{equation}
\end{widetext}
Inverting to the space-time domain $ s\rightarrow t $ and $ k\rightarrow x_{0} $ similar to that used in the derivation of the forward equation, in the continuum limit, we get the backward tempered fractional Feynman-Kac equation
\begin{widetext}
\begin{equation} \label{1stBTEq}
\frac{\partial}{\partial t}G_{x_{0}}(p,t)=\left[\lambda^{\alpha}D_{t}^{1-\alpha,\lambda}-\lambda\right]\left[G_{x_{0}}(p,t)-e^{-pU(x_{0})t}\right]
-pU(x_{0})G_{x_{0}}(p,t)+K_{\alpha}D_{t}^{1-\alpha,\lambda}\frac{\partial^{2}}{\partial x_{0}^{2}}G_{x_{0}}(p,t).
\end{equation}
\end{widetext}
The initial condition is $ G_{x_{0}}(A,t=0)=\delta(A) $, or in Laplace space $ G_{x_{0}}(p,t=0)=1 $, where $ \delta(\cdot) $ is also the Dirac delta function. The symbol $ D_{t}^{1-\alpha,\lambda} $ is the tempered fractional substantial derivative, defined as Eq. (\ref{temperedFSD}). Notice that here, this operator appears to the left of the Laplacian $ \frac{\partial^{2}}{\partial x_{0}^{2}} $ in Eq. (\ref{1stBTEq}), in contrast to the forward equation (\ref{1stFTEq}). 
When $\lambda=0$, Eq. (\ref{1stBTEq}) turns to the backward fractional Feynman-Kac equation \cite{Turgeman:00}:
$$ \frac{\partial}{\partial t}G_{x_{0}}(p,t)=K_{\alpha}D_{t}^{1-\alpha}\frac{\partial^{2}}{\partial x_{0}^{2}}G_{x_{0}}(p,t)-pU(x_{0})G_{x_{0}}(p,t). $$

\subsection{A one-dimensional lattice random walk with forces}
This subsection still considers the CTRW on lattice but with forces, which means the probabilities of jumping left ($L(x)$) and right ($R(x)$) are no longer equal. Assume the system is coupled to a heat bath at temperature $T$ and detailed balance, i.e., $L(x) \exp \left[ -\frac{V(x)}{k_B T} \right]=R(x-a) \exp \left[-\frac{V(x-a)}{k_B T} \right]$, where $a$ is the spacing of the lattice. For small $a$, expanding $R(x)$, $L(x)$, and the exponential function leads to
$$R(x)\simeq \frac{1}{2}\left[ 1+\frac{a F(x)}{2k_BT} \right] ,  ~ L(x)\simeq \frac{1}{2}\left[ 1-\frac{a F(x)}{2k_BT} \right],$$ where $F(x)=-V^\prime(x)$ \cite{Carmi:01}.

{\em Derivation of the forward tempered fractional Feynman-Kac equation with forces.} --- Using $ \cos(ka)\simeq 1-\frac{a^{2}k^{2}}{2}$ and $\sin(ka) \simeq ka $ for the long wavelength $ k\rightarrow 0$ and following Eq. (18) in \cite{Carmi:01},
\begin{widetext}\begin{equation} \label{eq13}
G(k,p,s)\simeq\frac{1-\hat{\psi}[s+pU(-i\frac{\partial}{\partial
k}),\lambda]}{s+pU(-i\frac{\partial}{\partial k})}\cdot
\frac{1}{1-\left[1-\frac{a^{2}k^{2}}{2}
+i(ka)\frac{aF(-i\frac{\partial}{\partial K})}{2k_{b}T}\right
]\hat{\psi}[s+pU(-i\frac{\partial}{\partial k}),\lambda]}.
\end{equation}
Substituting $\hat{\psi}(s,\lambda)$ Eq. (\ref{LaptemperedWTPDF}) into Eq. (\ref{eq13}) and rearranging the equation, we obtain
\begin{equation}
\begin{split}
\frac{a^{2}}{2B_{\alpha}}\left\{k^{2} -ik\frac{F\left(-i\frac{\partial}{\partial k}\right)}{k_{b}T}\right\}\left[\lambda+s+pU\left(-i\frac{\partial}{\partial k}\right)\right]^{1-\alpha}G(k,p,s)
 +\left[\lambda+s+pU\left(-i\frac{\partial}{\partial k}\right)\right]G(k,p,s)
 \\
 -\lambda^{\alpha}\left[\lambda+s+pU\left(-i\frac{\partial}{\partial k}\right)\right]^{1-\alpha}G(k,p,s)-1
=\frac{\lambda-\lambda^{\alpha}[\lambda+s+pU(-i\frac{\partial}{\partial k})]^{1-\alpha}}{s+pU(-i\frac{\partial}{\partial k})}.
\end{split}
\end{equation}
Inverting $ k\rightarrow x $, $ s\rightarrow t $, then
\begin{equation} \label{2ndFTEq}
\begin{array}{c}
\frac{\partial}{\partial t}G(x,p,t)=
\left[\lambda^{\alpha}D_{t}^{1-\alpha,\lambda}-\lambda\right]\left[G(x,p,t)-e^{-pU(x)t}\delta(x)\right]-pU(x)G(x,p,t)
+K_{\alpha}\left[\frac{\partial^{2}}{\partial x^{2}}-\frac{\partial}{\partial x}\frac{F(x)}{k_{b}T}\right]D_{t}^{1-\alpha,\lambda}G(x,p,t).
\end{array}
\end{equation}
Similarly, we can obtain another equation as follows,
\begin{equation} \label{a2stFTEq}
 \frac{\partial}{\partial t}G(x,p,t)+pU(x)G(x,p,t)=  \\ K_{\alpha}\left[\frac{\partial^{2}}{\partial x^{2}}-\frac{\partial}{\partial x}\frac{F(x)}{k_{b}T}\right]\left[\frac{\partial}{\partial t}+pU(x)\right]\int_{0}^{t}K(t-\tau,\lambda)e^{-pU(x)(t-\tau)}G(x,p,\tau)d\tau.
\end{equation}
If $F(x)=0$, then Eq. (\ref{2ndFTEq}) is the same as Eq. (\ref{1stFTEq}).
If $\lambda=0$, Eq. (\ref{2ndFTEq}) becomes the same as Eq. (22) given in \cite{Carmi:01},
$$ \frac{\partial}{\partial t}G(x,p,t)=K_{\alpha}\left[\frac{\partial^{2}}{\partial x^{2}}-\frac{\partial}{\partial x}\frac{F(x)}{k_{b}T}\right]D_{t}^{1-\alpha}G(x,p,t)-pU(x)G(x,p,t). $$
\end{widetext}

{\em Derivation of the backward tempered fractional Feynman-Kac equation with forces.} --- As mentioned in the above subsection, if we are just interested in the distribution of the functional $A$, the backward equation should be useful and convenient. For $G_{k}(p,s)$, the following formal equation holds \cite{Carmi:01},
\begin{widetext}
\begin{equation} \label{eq16}
 G_{k}(p,s)\simeq \hat{W}\left[pU\left(-i\frac{\partial}{\partial k}\right)+s,\lambda\right]\delta(k)+\hat{\psi}\left[pU\left(-i\frac{\partial}{\partial k}\right)+s,\lambda\right]
 \cdot \left[\cos(ka)-\frac{aF(-i\frac{\partial}{\partial k})}{2k_{b}T}i\sin(ka)\right]G_{k}(p,s).
\end{equation}
Substituting $ \hat{W}(s)=(1-\hat{\psi}(s,\lambda))/s $ and $ \hat{\psi}(s,\lambda) $ (given in Eq. (\ref{LaptemperedWTPDF})) into Eq. (\ref{eq16}), and using $ \cos(ka)\simeq 1-\frac{k^{2}a^{2}}{2} $ and $ \sin(ka)\simeq ka $ as $ ak\rightarrow 0 $ and small $s$ ($B_{\alpha}\rightarrow 0$) approximation, after some rearrangements, we have
\begin{equation}
\begin{split}
 \frac{a^{2}}{2B_{\alpha}}&\left[\lambda+pU\left(-i\frac{\partial}{\partial k}\right)+s\right]^{1-\alpha}\left[k^2+ \frac{F(-i\frac{\partial}{\partial k})}{k_{b}T}(ik)\right]G_{k}(p,s)+\left[\lambda+pU\left(-i\frac{\partial}{\partial k}\right)+s\right]G_{k}(p,s) \\
& -\lambda^{\alpha}\left[\lambda+pU\left(-i\frac{\partial}{\partial k}\right)+s\right]^{1-\alpha}G_{k}(p,s)-\delta(k)
 =\frac{\lambda-\lambda^{\alpha}[\lambda+pU(-i\frac{\partial}{\partial k})+s]^{1-\alpha}}{pU(-i\frac{\partial}{\partial k})+s}\delta(k).
\end{split}
\end{equation}
Taking inversion in the above equation, $ k\rightarrow x_{0} $ and $s\rightarrow t$, we get
\begin{equation} \label{2ndBTEq}
\begin{array}{c}
\frac{\partial}{\partial t}G_{x_{0}}(p,t)=\left[\lambda^{\alpha}D_{t}^{1-\alpha,\lambda}-\lambda\right]\left[G_{x_{0}}(p,t)-e^{-pU(x_{0})t}\right]
-pU(x_{0})G_{x_{0}}(p,t)+K_{\alpha}D_{t}^{1-\alpha,\lambda}\left[\frac{\partial^{2}}{\partial x_{0}^{2}}+\frac{F(x_{0})}{k_{b}T}\frac{\partial}{\partial x_{0}}\right]G_{x_{0}}(p,t).
\end{array}
\end{equation}
If $ F(x)=0 $, then Eq. (\ref{2ndBTEq}) is exactly the same as Eq. (\ref{1stBTEq}).
For $ \lambda=0 $, Eq. (\ref{2ndBTEq}) reduces to the one given in \cite{Carmi:01}:
$$ \frac{\partial}{\partial t}G_{x_{0}}(p,t)=K_{\alpha}D_{t}^{1-\alpha}\left[\frac{\partial^{2}}{\partial x_{0}^{2}}+\frac{F(x_{0})}{k_{b}T}\frac{\partial}{\partial x_{0}}\right]G_{x_{0}}(p,t)-pU(x_{0})G_{x_{0}}(p,t). $$
\end{widetext}

\subsection{Tempered CTRW with power-law jump length distribution}
Instead of discussing the tempered CTRW on a lattice, we further analyze the tempered CTRW with a power-law jump length distribution, $ \eta(x)\simeq|x|^{-1-\beta}$, $0<\beta<2$, and the Fourier transform of $\eta(x)$ is \cite{Metzler:00}
\begin{equation}\label{eta_k}
\eta(k)=\exp(-C_\beta|k|^\beta) \simeq 1-C_{\beta}|k|^{\beta}.
\end{equation}
Tempering of jump length will be considered in the next subsection.

{\em Derivation of the forward tempered fractional Feynman-Kac equation with power-law jump length distribution.} --- From the CTRW model, the main Eq.(\ref{GkpsLatt}) is modified according
\begin{widetext}
\begin{equation} \label{eq20}
G(k,p,s)=\frac{1-\hat{\psi}[s+pU(-i\frac{\partial}{\partial k}),\lambda]}{s+pU(-i\frac{\partial}{\partial k})} \frac{1}{1-\eta(k)\hat{\psi}[s+pU(-i\frac{\partial}{\partial k}),\lambda]}.
\end{equation}
Compared with Eq. (\ref{GkpsLatt}) where random walk is on a lattice, hence Fourier transform of jump length PDF was $\cos(ka)$, now we replace it with the more general form $\eta(k)$.
Substituting the approximation of $\eta(k)$ (given in Eq. (\ref{eta_k})) and $ \hat{\psi}(s,\lambda) $ (given in Eq. (\ref{LaptemperedWTPDF})) into Eq. (\ref{eq20}) leads to
\begin{equation} \label{eq21}
\begin{split}
 G(k,p,s)\simeq\frac{B_{\alpha}[\lambda+s+pU(-i\frac{\partial}{\partial k})]^{\alpha}-B_{\alpha} \lambda^{\alpha}}{s+pU(-i\frac{\partial}{\partial k})}
          \cdot\frac{1}{1-(1-C_{\beta}|k|^{\beta})[1-B_{\alpha}(\lambda+s+pU(-i\frac{\partial}{\partial k}))^{\alpha}+B_{\alpha} \lambda^{\alpha}]}.
\end{split}
\end{equation}
Rearranging Eq. (\ref{eq21}) and taking $k\rightarrow 0$, we obtain the following equation
\begin{equation}
\begin{split}
& \frac{C_{\beta}}{B_{\alpha}}|k|^{\beta}\left[\lambda+s+pU\left(-i\frac{\partial}{\partial k}\right)\right]^{1-\alpha}G(k,p,s)
 +\left[\lambda+s+pU\left(-i\frac{\partial}{\partial k}\right)\right]G(k,p,s)-1 \\
& -\lambda^{\alpha}\left[\lambda+s+pU\left(-i\frac{\partial}{\partial k}\right)\right]^{1-\alpha}G(k,p,s)
 =\frac{\lambda-\lambda^{\alpha}[\lambda+s+pU(-i\frac{\partial}{\partial k})]^{1-\alpha}}{s+pU(-i\frac{\partial}{\partial k})}.
\end{split}
\end{equation}
Taking $k\rightarrow x, s\rightarrow t$ in the above equation results in the forward Feynman-Kac equation
\begin{equation} \label{3rdFTEq}
\frac{\partial}{\partial t}G(x,p,t)=
\left[\lambda^{\alpha}D_{t}^{1-\alpha,\lambda}-\lambda\right]\left[G(x,p,t)-e^{-pU(x)t}\delta(x)\right]-pU(x)G(x,p,t)
+\frac{C_{\beta}}{B_{\alpha}}\nabla_{x}^{\beta}D_{t}^{1-\alpha,\lambda}G(x,p,t),
\end{equation}
\end{widetext}
where the Riesz spatial fractional derivative operator $ \nabla_{x}^{\beta} $ and the fractional Laplacian operator $ -(-\Delta_{x})^{\beta/2} $ are equivalent \cite{Yang:00}. In Fourier $ x\rightarrow k $ space $ \nabla_{x}^{\beta}\rightarrow -|k|^{\beta} $ \cite{Carmi:00}; and in $ x $ space,
$$ \nabla_{x}^{\beta}f(x)=-\frac{1}{2\cos\frac{\beta\pi}{2}}
\left[_{-\infty}D_{x}^{\beta}f(x)+\,_{x}D_{+\infty}^{\beta}f(x)\right],$$
where $(n-1 <\beta< n)$
\begin{equation}
_{-\infty}D_{x}^{\beta}f(x)=\frac{1}{\Gamma(n-\beta)}\frac{d^{n}}{dx^{n}}\int_{-\infty}^{x}\frac{f(\xi)}{(x-\xi)^{\beta+1-n}}d\xi,
\end{equation}
\begin{equation}
_{x} D_{+\infty}^{\beta}f(x)=\frac{(-1)^{n}}{\Gamma(n-\beta)}
\frac{d^{n}}{dx^{n}}\int_{x}^{+\infty}\frac{f(\xi)}{(\xi-x)^{\beta+1-n}}d\xi.
\end{equation}
And the symbol $ D_{t}^{1-\alpha,\lambda} $ is defined as before. If taking $\beta=2$, Eq. (\ref{3rdFTEq}) reduces to Eq. (\ref{1stFTEq}); letting $\lambda=0$ leads to
 $\frac{\partial}{\partial t}G(x,p,t)=\frac{C_{\beta}}{B_{\alpha}}\nabla_{x}^{\beta}D_{t}^{1-\alpha}G(x,p,t)-pU(x)G(x,p,t)$,
 which is the same as the one obtained in \cite{Carmi:00}.

{\em Derivation of the backward tempered Feynman-Kac equation with power-law jump length distribution.} --- Following \cite{Carmi:10} and replacing $\cos(ka)$ with $\eta(k)$ in Eq. (\ref{beq}) corresponded to the general jump lengths, there exists
\begin{widetext}
\begin{equation} \label{eq24}
\begin{split}
  G_{k}(p,s)  =\hat{\psi}\left[pU\left(-i\frac{\partial}{\partial k}\right)+s,\lambda\right]\eta(k)G_{k}(p,s)+\hat{W}\left[pU\left(-i\frac{\partial}{\partial
  k}\right)+s,\lambda\right]\delta(k).
\end{split}
\end{equation}
Substituting the approximation of $\eta(k)$ (given in Eq. (\ref{eta_k})),  $
\hat{W}(s,\lambda)=(1-\hat{\psi}(s,\lambda))/s $, and $ \hat{\psi}(s,\lambda) $ (given in
Eq. (\ref{LaptemperedWTPDF})) into Eq. (\ref{eq24}), taking $
k\rightarrow 0 $ and rearranging the terms, we find
\begin{equation}
\begin{split}
\left[\lambda+pU\left(-i\frac{\partial}{\partial k}\right)+s \right]^{1-\alpha}\frac{C_{\beta}}{B_{\alpha}}|k|^{\beta}G_{k}(p,s)
   +\left\{\left[\lambda+pU\left(-i\frac{\partial}{\partial k}\right)+s\right]
   -\lambda^{\alpha}\left[\lambda+pU\left(-i\frac{\partial}{\partial k}\right)+s\right]^{1-\alpha}\right\}G_{k}(p,s)
   \\
   -\delta(k)=\frac{\lambda-\lambda^{\alpha}[\lambda+pU(-i\frac{\partial}{\partial k})+s]^{1-\alpha}}{pU(-i\frac{\partial}{\partial k})+s}\delta(k).
\end{split}
\end{equation}
Taking the inverse transforms $ k\rightarrow x_{0}, s\rightarrow t $, we get
\begin{equation} \label{3rdBTEq}
\frac{\partial}{\partial t}G_{x_{0}}(p,t)=\left[\lambda^{\alpha}D_{t}^{1-\alpha,\lambda}-\lambda\right]\left[G_{x_{0}}(p,t)-e^{-pU(x_{0})t}\right]
-pU(x_{0})G_{x_{0}}(p,t)+\frac{C_{\beta}}{B_{\alpha}}D_{t}^{1-\alpha,\lambda}\nabla_{x_{0}}^{\beta}G_{x_{0}}(p,t).
\end{equation}
If taking $\beta=2$, Eq. (\ref{3rdBTEq}) reduces to Eq. (\ref{1stBTEq}); while $\lambda=0$, it reduces to Eq. (21)
in \cite{Carmi:10} as expected.

\end{widetext}
\subsection{Tempered CTRW with tempered power-law jump length
distribution}\label{SubSecD}

Now we are going to discuss the tempered power-law jump length distribution $ \eta(x)\simeq \frac{A_\beta}{|\Gamma(-\beta)|} e^{-\gamma |x|}|x|^{-\beta-1} $, where $ 0<\gamma,\, 0<\beta<2$; and the asymptotic form (with small $k$) of the Fourier transform of $\eta(x)$ is \cite{Nakao:11}
\begin{equation}\label{eta_k2}
\eta(k) \simeq 1-A_{\beta}^{\theta}(\gamma^2+k^2)^{\beta/2}+2A_{\beta}\gamma^{\beta},
\end{equation}
where $ \theta=\arg(\gamma+ik) $, $
A_{\beta}^{\theta}=2A_{\beta}\cos(\beta\theta) $ .

{\em Derivation of the forward tempered fractional Feynman-Kac
equation with tempered power-law jump length distribution.} ---
Similar to the above analysis, substituting the approximation of $\eta(k)$ (given in Eq.
(\ref{eta_k2})) and  $ \hat{\psi}(s,\lambda) $ (given in Eq.
(\ref{LaptemperedWTPDF})) into Eq. (\ref{eq20}), we get
\begin{widetext}
\begin{equation}\label{eq28}
  G(k,p,s)\simeq\frac{B_{\alpha}[\lambda+s+pU(-i\frac{\partial}{ \partial k})]^\alpha-B_{\alpha}\lambda^\alpha}{s+pU(-i\frac{\partial}{ \partial k})}\cdot\frac{1}{1-[1-A_{\beta}^{\theta}(\gamma^2+k^2)^{\beta/2}+2A_{\beta}\gamma^{\beta}]\{1-B_{\alpha}[\lambda+s+pU(-i\frac{\partial}{\partial k})]^{\alpha}+B_{\alpha}\lambda^{\alpha}\}}.
\end{equation}

Taking $ k\rightarrow 0 $ makes $ A_{\beta}^{\theta}\rightarrow
2A_{\beta} $. Rearranging Eq. (\ref{eq28}), we have
\begin{equation}
\begin{split}
 \left\{ K_{\alpha,\beta}  (\gamma^2+k^2)^{\beta/2}\left[\lambda+s+pU\left(-i\frac{\partial}{\partial k}\right)\right]^{1-\alpha}-
  K_{\alpha,\beta}\gamma^\beta\left[\lambda+s+pU\left(-i\frac{\partial}{\partial k}\right)\right]^{1-\alpha}\right\}G(k,p,s)
  \\
  +\left[\lambda+s+pU\left(-i\frac{\partial}{\partial k}\right)\right]G(k,p,s)
  -\lambda^{\alpha}\left[\lambda+s+pU\left(-i\frac{\partial}{\partial k}\right)\right]^{1-\alpha}G(k,p,s)-1
\\
 =\frac{\lambda}{s+pU(-i\frac{\partial}{\partial k})}-\frac{\lambda^\alpha[\lambda+s+pU(-i\frac{\partial}{\partial k})]^{1-\alpha}}{s+pU(-i\frac{\partial}{\partial
 k})},
\end{split}
\end{equation}
where 
 $
K_{\alpha,\beta}=\frac{2A_{\beta}}{B_{\alpha}} $. Taking the
inversion transforms $ k\rightarrow x $ and $ s\rightarrow t $
results in
\begin{equation}\label{4thFTEq}
\frac{\partial}{\partial t}G(x,p,t)=
\left[\lambda^{\alpha}D_{t}^{1-\alpha,\lambda}-\lambda\right]\left[G(x,p,t)-e^{-pU(x)t}\delta(x)\right]-pU(x)G(x,p,t)
+K_{\alpha,\beta}\left(\nabla_{x}^{\beta,\gamma}+\gamma^{\beta}\right)D_{t}^{1-\alpha,\lambda}G(x,p,t).
\end{equation}
The tempered fractional Riesz derivative (TFRD) operator $ \nabla_{x}^{\beta,\gamma} $ is defined in Fourier $
x\rightarrow k $ space as $ \nabla_{x}^{\beta,\gamma}\rightarrow
-(\gamma^2+k^2)^{\beta/2}$; and in $ x $ space, the operator is defined as (for more details, see Appendix B):
\begin{equation}
\nabla_{x}^{\beta,\gamma}f(x)=-\frac{1}{2\cos(\frac{\beta\pi}{2})}\left[_{-\infty}\mathbb{D}_{x}^{\beta,\gamma}f(x)
+\,_{x}\mathbb{D}_{+\infty}^{\beta,\gamma}f(x)\right].
\end{equation}
When $ \gamma=0 $, Eq.
(\ref{4thFTEq}) becomes Eq. (\ref{3rdFTEq}) as expected.

{\em Derivation of the backward tempered fractional Feynman-Kac
equation with tempered power-law jump length distribution.} --- Again
following \cite{Carmi:10} and inserting the approximation of $\eta(k)$ (given in Eq.
(\ref{eta_k2})), $ \hat{W}(s,\lambda)=(1-\hat{\psi}(s,\lambda))/s $, and $
\hat{\psi}(s,\lambda) $ (given in Eq. (\ref{LaptemperedWTPDF})) into Eq.
(\ref{eq24}), we have
\begin{equation}
\begin{split}
  G_{k}(p,s) =& \hat{\psi}\left[pU\left(-i\frac{\partial}{\partial k}\right)+s,\lambda\right]\eta(k)G_{k}(p,s)+\hat{W}\left[pU\left(-i\frac{\partial}{\partial k}+s\right),\lambda\right]\delta(k) \\
     \simeq& \left\{1-B_{\alpha}\left[\lambda+pU\left(-i\frac{\partial}{\partial k}\right)+s\right]^{\alpha}+B_{\alpha}\lambda^{\alpha}\right\}\left[1-A_{\beta}^{\theta}(\gamma^2+k^2)^{\beta/2}+2A_{\beta}\gamma^{\beta}\right]G_{k}(p,s)
    \\
    & +\frac{B_{\alpha}[\lambda+pU(-i\frac{\partial}{\partial k})+s]^{\alpha}-B_{\alpha}\lambda^{\beta}}{pU(-i\frac{\partial}{\partial k})+s}\delta(k).
\end{split}
\end{equation}
Letting $ k\rightarrow 0 $ makes $ A_{\beta}^{\theta}\rightarrow
2A_{\beta} $. Rearranging the last equation leads to
\begin{equation}
\begin{split}
    \left[\lambda+pU\left(-i\frac{\partial}{\partial k}\right)+s\right]^{1-\alpha}[K_{\alpha,\beta}^{\theta}(\gamma^2+k^2)^{\beta/2}-K_{\alpha,\beta}\gamma^{\beta}]G_{k}(p,s)
\\
    + \left\{\left[\lambda+pU\left(-i\frac{\partial}{\partial k}\right)+s\right]-\lambda^{\alpha}\left[\lambda+pU\left(-i\frac{\partial}{\partial k}\right)+s\right]^{1-\alpha}\right\}G_{k}(p,s) \\
     =\frac{\lambda \delta(k)}{pU(-i\frac{\partial}{\partial k})+s}-\frac{\lambda^\alpha[\lambda+pU(-i\frac{\partial}{\partial k})+s]^{1-\alpha}}{pU(-i\frac{\partial}{\partial
    k})+s}\delta(k)+\delta(k),
\end{split}
\end{equation}
where $
K_{\alpha,\beta}^{\theta}=\frac{A_{\beta}^{\theta}}{B_{\alpha}} $ ,
$ K_{\alpha,\beta}=\frac{2A_{\beta}}{B_{\alpha}} $. Taking the
inverse Laplace and Fourier transformations, $ k\rightarrow x $ and
$ s\rightarrow t $, there exists
\begin{equation}\label{4thBTEq}
\frac{\partial}{\partial x}G_{x_{0}}(p,t)
=\left[\lambda^{\alpha}D_{t}^{1-\alpha,\lambda}-\lambda\right]\left[G_{x_{0}}(p,t)-e^{-pU(x_{0})t}\right]-pU(x_{0})G_{x_{0}}(p,t)
+K_{\alpha,\beta}D_{t}^{1-\alpha,\lambda}\left(\nabla_{x_{0}}^{\beta,\gamma}+\gamma^{\beta}\right)G_{x_{0}}(p,t).
\end{equation}
Notice that $ D_{t}^{1-\alpha,\lambda} $ is on the left of $
\nabla_{x_{0}}^{\beta,\gamma} $ in Eq. (\ref{4thBTEq}), in contrast
to the forward equation Eq. (\ref{4thFTEq}). When $\gamma=0 $, Eq.
(\ref{4thBTEq}) reduces to Eq. (\ref{3rdBTEq}) as expected.
\end{widetext}
%
\section{Solutions to the Derived Equations}
In this section, we present the distributions of four concrete functionals of the paths of particles performing tempered anomalous dynamics.

\subsection{Occupation time in half-space}
The occupation time of a particle in half-space is widely used in physics \cite{Stefani:00,Majumdar:00,Barkai:04} and mathematics \cite{Watanabe:00}. Define the occupation time in $ x>0 $ as $ T^{+}=A=\int_{0}^{t}\Theta[x(\tau)]d\tau $, i.e., $ U(x)=\Theta(x)=1 $ for $ x\geq0 $ and is zero otherwise. For example, for Brownian motion the PDF of $T^{+}$ is the famous Arcsine distribution. In order to find the PDF of the occupation time, here we consider the backward fractional Feynman-Kac equation (\ref{1stBTEq}) with regular jump length in Laplace $s$ space:
\begin{widetext}
\begin{equation}
-K_{\alpha}(\lambda+s)^{1-\alpha}\frac{\partial^{2}}{\partial x_{0}^{2}}G_{x_{0}}(p,s)+(\lambda+s)G_{x_{0}}(p,s)-1=\lambda^{\alpha}(\lambda+s)^{1-\alpha}G_{x_{0}}(p,s)+\frac{\lambda-\lambda^{\alpha}(\lambda+s)^{1-\alpha}}{s}, x_{0}<0.
\end{equation}
\begin{equation}
-K_{\alpha}(\lambda+s+p)^{1-\alpha}\frac{\partial^{2}}{\partial x_{0}^{2}}G_{x_{0}}(p,s)+(\lambda+s+p)G_{x_{0}}(p,s)-1=\lambda^{\alpha}(\lambda+s+p)^{1-\alpha}G_{x_{0}}(p,s)+\frac{\lambda-\lambda^{\alpha}(\lambda+s+p)^{1-\alpha}}{s+p},
x_{0}>0.
\end{equation}
\end{widetext}
Hence here the tempering is in time only and $K_{\alpha}$ is given in Eq. (\ref{Kalpha}). Rewriting the above equations leads to
\begin{equation}G_{x_{0}}(p,s)= \left\{
\begin{array}{ll}
K_{\alpha}\frac{1}{(\lambda+s)^{\alpha}-\lambda^{\alpha}}\frac{\partial^{2}}{\partial x_{0}^{2}}G_{x_{0}}(p,s)+\frac{1}{s}, & x_{0}<0; \\
\\
K_{\alpha}\frac{1}{(\lambda+s+p)^{\alpha}-\lambda^{\alpha}}\frac{\partial^{2}}{\partial x_{0}^{2}}G_{x_{0}}(p,s)+\frac{1}{s+p}, & x_{0}>0.
\end{array}
\right.
\end{equation}
They both are second order, ordinary differential equations in $ x_{0} $. Solving the equations in each half-space individually, requiring that $G_{x_{0}}(p,s)$ is finite for $ |x_{0}|\rightarrow \infty $,
\begin{equation} \label{eq40}
G_{x_{0}}(p,s)= \left\{
\begin{array}{l}
C_{0}\exp\left(x_{0}\sqrt{\frac{(\lambda+s)^{\alpha}-\lambda^{\alpha}}{K_{\alpha}}}\right)+\frac{1}{s}, ~~ x_{0}<0; \\
\\
C_{1}\exp\left(-x_{0}\sqrt{\frac{(\lambda+s)^{\alpha}-\lambda^{\alpha}}{K_{\alpha}}}\right)+\frac{1}{s+p},  \\ ~~~~~~~~~~~~~~~~~~~~~~~~~~~~~~~~~~~~~~~~~~ x_{0}>0.
\end{array}
\right.
\end{equation}
The particle can never arrive at $ x>0 $ for $ x_{0}\rightarrow -\infty $; thus $ G_{x_{0}}(T^{+},t)=\delta(T^{+}) $ and $ G_{x_{0}}(p,s)=\frac{1}{s} $, in conformity to Eq. (\ref{eq40}). Likewise, for $ x_{0}\rightarrow +\infty $, the particle is never at $ x<0 $ and thus $ G_{x_{0}}(T^{+},t)=\delta(T^{+}-t) $ and $ G_{x_{0}}(p,s)=\frac{1}{s+p} $, as expected in Eq. (\ref{eq40}). Then demanding that $ G_{x_{0}}(p,s) $ and its first derivative are continuous at $ x_{0}=0 $, yields a pair of equations about $ C_{0}, C_{1} $:
\begin{equation} \left\{
  \begin{array}{ll}
    C_{0}+\frac{1}{s}=C_{1}+\frac{1}{s+p} \\
\\
    C_{0}\sqrt{(\lambda+s)^{\alpha}-\lambda^{\alpha}}=-C_{1}\sqrt{(\lambda+s+p)^{\alpha}-\lambda^{\alpha}}.
  \end{array}
\right.
\end{equation}
By solving these equations, we get
\begin{equation} \label{eq42}
 \left\{
\begin{array}{ll}
C_{0}=-\frac{p\sqrt{(\lambda+s+p)^{\alpha}-\lambda^{\alpha}}}{s(s+p)(\sqrt{(\lambda+s+p)^{\alpha}-\lambda^{\alpha}}+\sqrt{(\lambda+s)^{\alpha}-\lambda^{\alpha}})}\\
\\
C_{1}=\frac{p\sqrt{(\lambda+s)^{\alpha}-\lambda^{\alpha}}}{s(s+p)(\sqrt{(\lambda+s+p)^{\alpha}-\lambda^{\alpha}}+\sqrt{(\lambda+s)^{\alpha}-\lambda^{\alpha}})}.
\end{array}
\right.
\end{equation}
Assume that the particle starts at $ x_{0}=0 $. Substituting $ x_{0}=0 $ in Eq. (\ref{eq40}), then $ G_{0}(p,s)=C_{0}+\frac{1}{s}=C_{1}+\frac{1}{s+p} $, i.e.,
\begin{equation}\label{eq43} G_{0}(p,s)=\frac{s\sqrt{(\lambda+s+p)^{\alpha}-\lambda^{\alpha}}+(s+p)\sqrt{(\lambda+s)^{\alpha}-\lambda^{\alpha}}}{s(s+p)(\sqrt{(\lambda+s+p)^{\alpha}-\lambda^{\alpha}}+\sqrt{(\lambda+s)^{\alpha}-\lambda^{\alpha}})}, \end{equation}
which describes the PDF of $ T^{+} $ and is valid for all times. However, it seems difficult to invert Eq. (\ref{eq43}) analytically. We'll soon analyse the moments of this equation in the following discussion. Specially, if $\alpha=1$, then $G_{0}(p,s)=s^{-1/2}(s+p)^{-1/2}$, this can be inverted to give the equilibrium PDF of $\varepsilon\equiv T^{+}/t$, or the occupation fraction, $$G(\varepsilon)=\frac{1}{\pi\sqrt{\varepsilon(1-\varepsilon)}},$$
which is the arcsine law of L\'{e}vy \cite{Majumdar:00}.

\subsection{First passage time}
\begin{figure}
  \centering
  \includegraphics[height=5cm,width=8cm]{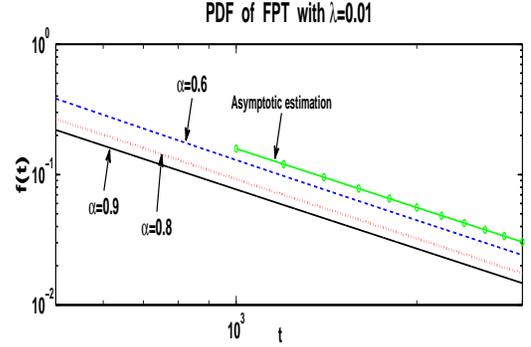}\\
  \caption{Behavior of $f(t)$ (i.e., Eq. (\ref{eq45})) with different values of the parameter $\alpha$, and lattice spacing $a=0.01$, starting point $b=0.05$ and diffusion constant $K_{\alpha}=1/2$.
  The solid dotted (green) line is the asymptotic estimation with the slope of -3/2 for long time, confirming that the standard Sparre Andersen scaling also holds for the tempered sub-diffusion.}\label{fig.1}
\end{figure}

\begin{figure}
  \centering
  \includegraphics[height=5cm,width=8cm]{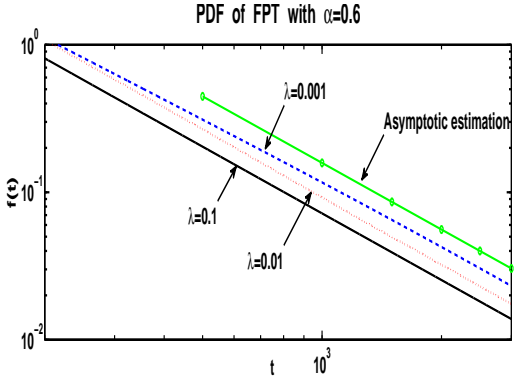}\\
  \caption{Behavior of $f(t)$ (i.e., Eq. (\ref{eq45})) with different values of the tempering parameter $\lambda$, and $a=0.01,b=0.05,K_{\alpha}=1/2$. The solid dotted (green) lines represent the asymptotic estimations with slope of -3/2 for long times.}\label{fig.2}
\end{figure}

\begin{figure}
  \centering
  \includegraphics[height=5cm,width=8cm]{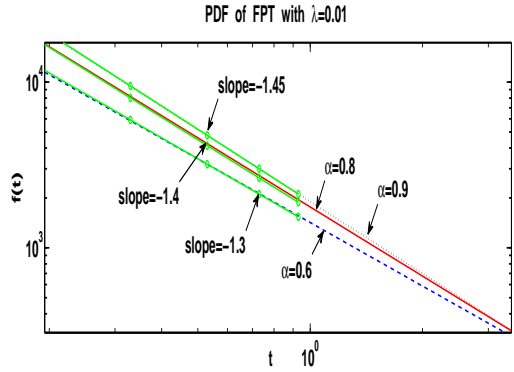}\\
  \caption{Behavior of $f(t)$ (i.e., Eq. (\ref{eq45})) with different values of the parameter $\alpha$, and $a=0.01,b=0.05,K_{\alpha}=1/2$. The solid dotted (green) lines represent the slope of $-(1+\alpha/2)$ for short but not too short times. }\label{fig.3}
\end{figure}

As well known, the first passage time (FPT) is defined as the time $ T_{f} $ it takes a particle starting at $ x_{0}=-b~(b>0) $ to hit $ x=0 $ for the first time \cite{Redner:00} and is widely applied in physics and other disciplines. A relationship between the distribution of first passage time and the occupation time functional was put forward by Kac \cite{Kac:00}:
$$ P_{r}\{T_{f}>t\}=P_{r}\{\displaystyle \max_{0 \leq \tau \leq t} x(\tau)<b \}=\displaystyle \lim_{p \rightarrow\infty}G_{x_{0}}(p,t),$$
where $ G_{x_{0}}(p,t) $ describes the Laplace transform of the PDF of functional $ T^{+}=\int_{0}^{t}\Theta[x(\tau)]d\tau $. The last equation is true for $ G_{x_{0}}(p,t)=\int_{0}^{\infty}e^{-pT^{+}}G_{x_{0}}(T^{+},t)dT^{+} $, and thus, if the particle has never passaged through $ x=0 $, we have $ T^{+}=0 $ and $ e^{-pT^{+}}=1 $, while otherwise, $ T^{+}>0 $, and for $ p \rightarrow \infty $, $ e^{-pT^{+}}=0 $. For $ x_{0}=-b $ and $ p \rightarrow \infty $, according to Eq. (\ref{eq40}) and (\ref{eq42}) of the previous subsection, we have
\begin{widetext}
\begin{equation}\label{eq44}\begin{split}
  \displaystyle \lim_{p \rightarrow\infty}G_{-b}(p,s)& = \frac{1}{s}-\displaystyle \lim_{p \rightarrow\infty}\frac{p\sqrt{(\lambda+s+p)^{\alpha}-\lambda^{\alpha}}}{s(s+p)(\sqrt{(\lambda+s+p)^{\alpha}-\lambda^{\alpha}}+
  \sqrt{(\lambda+s)^{\alpha}-\lambda^{\alpha}})}\exp\left(-b\sqrt{\frac{(\lambda+s)^{\alpha}-\lambda^{\alpha}}{K_{\alpha}}}\right) \\
                    & =\frac{1}{s}-\frac{1}{s}\exp\left(-b\sqrt{\frac{(\lambda+s)^{\alpha}-\lambda^{\alpha}}{K_{\alpha}}}\right).
                \end{split}
\end{equation}
\end{widetext}
In accordance with the definition of the first passage time, its PDF satisfies
$$ f(t)=\frac{\partial}{\partial t}(1-P_{r}\{T_{f}>t\})=-\frac{\partial}{\partial t} \displaystyle \lim_{p \rightarrow\infty}G_{-b}(p,t). $$
Hence, in Laplace space, we have
\begin{equation} \label{eq45}
\begin{split}
  f(s)& =-s\displaystyle \lim_{p \rightarrow\infty}G_{-b}(p,s)+1 \\
      & =\exp\left(-b\sqrt{\frac{(\lambda+s)^{\alpha}-\lambda^{\alpha}}{K_{\alpha}}}\right).
\end{split}
\end{equation}
The inversion of Eq. (\ref{eq45}) is done numerically \cite{Schneider:89} seeing Fig. \ref{fig.1}, Fig. \ref{fig.2} and Fig. \ref{fig.3}.

\begin{figure}
  \centering
  \includegraphics[height=5cm,width=8cm]{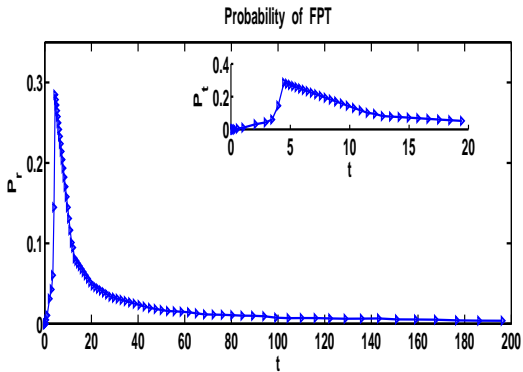}\\
  \caption{Simulations with $a=0.01,b=0.05,K_{\alpha}=1/2$ ended at $t=2\times10^{2}$ and included at least $6\times10^{4}$ trajectories for $\alpha=0.6$ and $\lambda=0.1$. The (blue) line with the mark of triangle is for the simulation result. One can see that $P_{r}$ increases from $0$ to a peak then decreases. }\label{fig.4}
\end{figure}

\begin{figure}
  \centering
  \includegraphics[height=5cm,width=8cm]{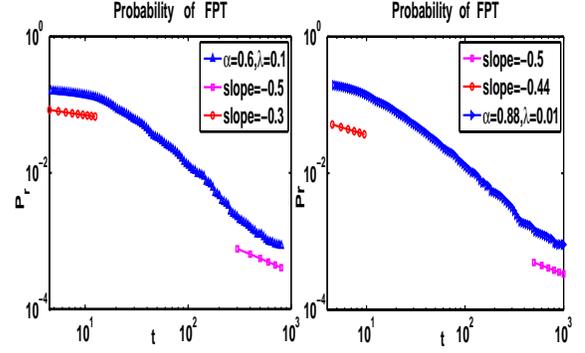}\\
  \caption{Behaviors of $P_{r}$ generated by $6\times10^{4}$ trajectories. For short but not too short times behaves as slope of $-\alpha/2$ (Fig. \ref{fig.3}) and large times behaves as slope of $-1/2$ (Eq. (\ref{Pr})) with $a=0.01,b=0.05,K_{\alpha}=1/2$. }\label{fig.5}
\end{figure}

Expanding Eq. (\ref{eq45}) in small $s$,
$$ f(s)\simeq 1-b\sqrt{\frac{\alpha\lambda^{\alpha-1}s}{K_{\alpha}}}. $$
Taking inverse Laplace transform for long times, $ s\rightarrow t $, we have
\begin{equation}
f(t)\simeq \frac{b}{|\Gamma(-\frac{1}{2})|}\sqrt{\frac{\alpha\lambda^{\alpha-1}}{K_{\alpha}}}t^{-\frac{3}{2}},
\end{equation}
for all $\alpha$, which coincides with the famous $ t^{-\frac{3}{2}} $ decay law of a one-dimensional random walk \cite{Redner:00,Schr:00} and decreases with increasing $\lambda $, being confirmed in Fig. \ref{fig.1} and Fig. \ref{fig.2}.
Hence,
\begin{equation}\label{Pr}
P_{r}\{T_{f}>t\}=\int_{t}^{\infty}f(T_{f})dT_{f} \simeq \frac{b}{\sqrt{\pi}}\sqrt{\frac{\alpha\lambda^{\alpha-1}}{K_{\alpha}}}t^{-1/2},
\end{equation}
the last equation is exactly the result given in \cite{Gajda:00} and is illustrated in Fig. \ref{fig.4} and Fig. \ref{fig.5}.

However, if $s\rightarrow \infty$, corresponding to small $t$, from Eq. (\ref{eq45}), we have
$$f(s)\simeq \exp\left(-\frac{b}{\sqrt{K_{\alpha}}}s^\frac{\alpha}{2}\right).$$
In $t$ space, the above equation tends to be the one-sided L\'{e}vy laws $L_{\alpha/2}(t)$. Hence $f(t)$ decays very fast to zero when $t\rightarrow0$ and behaviors as $t^{-1-\alpha/2}$ for short but not too short times, corresponded probability $P_{r}$ is illustrated in Fig. \ref{fig.4} and Fig. \ref{fig.5}, this is expected the particle cannot reach the origin when $t\rightarrow0$ and shows as L\'{e}vy behaviour for short but not too short times.

When $\lambda=0$, then waiting times show as power-law distributed. Eq. (\ref{eq45}) becomes
\begin{equation}\label{fs}
f(s)=\exp\left(-\frac{b}{\sqrt{K_{\alpha}}}s^\frac{\alpha}{2}\right).
\end{equation}
In $t$ space, Eq. (\ref{fs}) is the one-sided L\'{e}vy laws $L_{\alpha/2}(t)$. And then $f(t)$ decays very fast to zero when $t\rightarrow0$. For $t\rightarrow\infty$, $f(t)$ behaves as $t^{-(1+\alpha/2)}$, which is in agreement with the results given in \cite{Carmi:00,Barkai:00}, indicating that $\langle t\rangle$ is infinite for all $\alpha$.


\subsection{Maximal displacement}
Now we develop another application of Eq. (\ref{eq40}). The maximal displacement of a diffusing particle is a random variable, which has been studied in recent years \cite{Comtet:00,Schehr:00}. In order to get the distribution of this variable, we have $G_{x_{0}}(p,t)$ describes the functional $A=\int_{0}^{t}U[x(\tau)]d\tau$ with $U(x)=1$ for $x>0$, otherwise, $U(x)=0$. Let $ x_{m}\equiv \displaystyle \max_{0 \leq \tau \leq t} x(\tau) $; and then $ P_{r}\{x_{m}<b\}=\displaystyle \lim_{p\rightarrow \infty}G_{x_{0}}(p,t) $. From the last subsection we have, for $ x_{0}=-b $ (Eq. (\ref{eq44})),
$$ P_{r}\{x_{m}<b\}=\frac{1}{s}-\frac{1}{s}\exp\left(-b\sqrt{\frac{(\lambda+s)^{\alpha}-\lambda^{\alpha}}{K_{\alpha}}}\right). $$
Then the PDF of $ x_{m} $ is
$$ p(x_{m},s)=\frac{1}{s}\sqrt{\frac{(\lambda+s)^{\alpha}-\lambda^{\alpha}}{K_{\alpha}}}
\exp\left(-x_{m}\sqrt{\frac{(\lambda+s)^{\alpha}-\lambda^{\alpha}}{K_{\alpha}}}\right). $$
When $ \lambda=0 $, the above equation becomes
$$ p(x_{m},s)==\frac{1}{s}\sqrt{\frac{s^{\alpha}}{K_{\alpha}}}
\exp\left(-x_{m}\sqrt{\frac{s^{\alpha}}{K_{\alpha}}}\right). $$
Inverting $ s \rightarrow t $, $x_{m}>0$, we have \cite{Klafter:00}
$$ p(x_{m},t)=\sqrt{\frac{8}{\alpha^{2}K_{\alpha}}}\frac{t}{\left(x_{m}\sqrt{\frac{2}{K_{\alpha}}}\right)^{1+\frac{2}{\alpha}}}
L_{\frac{\alpha}{2}}\left[\frac{t}{\left(x_{m}\sqrt{\frac{2}{K_{\alpha}}}\right)^{\frac{2}{\alpha}}}\right].$$
The PDF is in agreement with the recent result of \cite{Schehr:00}, derived via a re-normalization group method.

\subsection{Fluctuations of occupation fraction}
In this subsection, we introduce a new variable $ \varepsilon $ which is defined as $ \varepsilon\equiv\frac{T^{+}}{t} $ or called occupation fraction \cite{Barkai:04,Godr:00}. As it is said in the above subsection, $ G_{0}(p,s) $ describing the occupation time functional in Eq. (\ref{eq43}) seems hard to be inverted analytically. Consequently, we would like to use the following method to calculate the first few moments
$$ \langle(T^{+})^{n}\rangle_{s}=(-1)^{n}\frac{\partial^{n}}{\partial p^{n}}G_{0}(p,s)|_{p=0}. $$
The first moment
$$ \langle T^{+}\rangle_{s}=-\frac{\partial}{\partial p}G_{0}(p,s)|_{p=0}=\frac{1}{2s^{2}}. $$
Performing the inversion, we have $ \langle T^{+}\rangle=\frac{t}{2} $ or $ \langle\varepsilon\rangle_{t}=\frac{1}{2} $ being the same as the case of $ \lambda=0 $ \cite{Carmi:00}. Namely, exponential tempering has no influence on the first moment of the occupation time as expected from symmetry. For the second moment,
\begin{equation} \label{eq46}
\langle (T^{+})^{2}\rangle_{s}=\frac{\partial^{2}}{\partial p^{2}}G_{0}(p,s)|_{p=0}=\frac{1}{s^{3}}-\frac{\alpha(s+\lambda)^{\alpha-1}}{4s^{2}[(s+\lambda)^{\alpha}-\lambda^{\alpha}]}.
\end{equation}
Inverting Eq. (\ref{eq46}), we get \cite{Grad:00}
\begin{equation} \label{eq47}
\langle (T^{+})^{2}\rangle\simeq \frac{t^{2}}{2}-\frac{\alpha}{4}t\ast e^{-\lambda t}E_{\alpha,1}[\lambda^{\alpha}t^{\alpha}],
\end{equation}
where the symbol $ \ast $ describes the convolution operator $ f(t)\ast g(t)=\int_{0}^{t}f(t-\tau)g(\tau)d\tau $; and we used the Laplace transform relation \cite{Pod:00}
$$ \int_{0}^{\infty}e^{-st}E_{\alpha,1}(at^{\alpha})dt=\frac{s^{\alpha-1}}{s^{\alpha}-a}, $$
and $ E_{\alpha,1}(z) $ is the Mittag-Leffler function, defined as
$$ E_{\alpha,1}(z)=\displaystyle \sum_{n=0}^{\infty}\frac{z^{n}}{\Gamma(1+\alpha n)}. $$

\textbf{D.1} As $ s\rightarrow \infty $, i.e., $t\rightarrow 0$, $ \lambda $ can be ignored in Eq. (\ref{eq46}), then we obtain $ \langle (T^{+})^{2}\rangle_{s}\simeq \frac{4-\alpha}{4s^{3}} $. Hence, $ \langle (T^{+})^{2}\rangle\simeq \frac{4-\alpha}{8}t^{2} $.
In fact, this result can also be derived in a different method.
When $ s\rightarrow \infty $, namely, $ t\rightarrow 0 $, we have, $ E_{\alpha,1}(\lambda^{\alpha}t^{\alpha})\simeq 1+\frac{\lambda^{\alpha}t^{\alpha}}{\Gamma(1+\alpha)}$. Hence we set
\begin{equation}
\begin{split}
    & t \ast e^{-\lambda t}E_{\alpha,1}(\lambda^{\alpha}t^{\alpha})
     \\
     \simeq & \int_{0}^{t}(t-\tau)e^{-\lambda \tau}\left[1+\frac{\lambda^{\alpha}\tau^{\alpha}}{\Gamma(1+\alpha)}\right]d\tau \\
     = &\frac{e^{-\lambda t}+\lambda t-1}{\lambda^{2}}+\frac{e^{-\lambda t}t^{\alpha+1}\lambda^{\alpha-1}}{\Gamma(\alpha+1)} \\
    &  +\frac{t\lambda^{\alpha}-(\alpha+1)\lambda^{\alpha-1}}{\Gamma(\alpha+1)}\int_{0}^{t}e^{-\lambda \tau}\tau^{\alpha}d\tau.
   \end{split}
\end{equation}
As $ t\rightarrow 0 $, the second and the third terms are zero, meanwhile, $ e^{-\lambda t}\simeq 1-\lambda t+\frac{\lambda^{2} t^{2}}{2} $. Therefore,
$$ t\ast e^{-\lambda t}E_{\alpha,1}(\lambda^{\alpha}t^{\alpha})\simeq \frac{t^{2}}{2}. $$
Substituting the above result into Eq. (\ref{eq47}), we get the same consequence
$$ \langle (T^{+})^{2}\rangle\simeq \frac{t^{2}}{2}-\frac{\alpha}{8}t^{2}=\frac{4-\alpha}{8}t^{2}. $$
Dividing by $ t^{2} $, we obtain the fluctuations of the occupation fraction, $ \langle(\bigtriangleup\varepsilon)^{2}\rangle_{t}=\langle\varepsilon^{2}\rangle_{t}-\langle\varepsilon\rangle_{t}^{2} $,
$$ \langle(\bigtriangleup\varepsilon)^{2}\rangle_{t}\simeq \frac{1-\alpha/2}{4}. $$
This is the expected result \cite{Carmi:01}. The reason is that for short times ($ \lambda $ has no effect on the process) the PDF Eq. (\ref{eq43}) is Lamperti's with index $ \alpha/2 $.\\

\textbf{D.2} As $s\rightarrow 0$, i.e., $ t\rightarrow \infty$, expanding Eq. (\ref{eq46}) in small $s$, we have
$$\langle(T^{+})^{2}\rangle_{s} \simeq \frac{1}{s^{3}}-\frac{1}{4s^{3}}=\frac{3}{4s^{3}}.$$
Taking inverse Laplace transform of the last equation, we find
$$\langle(T^{+})^{2}\rangle \simeq \frac{3}{8}t^{2}.$$
Then
$$ \langle(\bigtriangleup\varepsilon)^{2}\rangle_{t}\simeq \frac{1}{8}.$$
For $ 0<\alpha<1,~ t\rightarrow \infty $, $ \langle(\bigtriangleup\varepsilon)^{2}\rangle_{t} $ is always positive and is the same as the result we obtained in $\textbf{D.1}$ when $\alpha=1$.

\section{Fluctuations of the time-averaged position}
We analyze the time-averaged position, $ \overline{x}(t)=\int_{0}^{t}x(\tau)d\tau/t=\frac{A}{t} $, or in other words, taking $ U(x)=x $, for a tempered subdiffusive particle in a harmonic potential, $V(x)=\frac{m\omega^{2}x^{2}}{2}$. Then we treat the problem of the fluctuations of the time-average of position,
\begin{equation}\label{fluctuaions}
\langle(\triangle\overline{x})^{2}\rangle_{t}=\langle A^{2}\rangle/t^{2},
\end{equation}
where $x_{0}=0$ is assumed such that $ \langle \overline{x}\rangle=0 $ at all times due to symmetry. Hence, define the Fokker-Planck operator as $ L_{fp}=K_{\alpha}[\frac{\partial^{2}}{\partial x^{2}}+\frac{\partial}{\partial x}\frac{m\omega^{2}x}{k_{b}T})] $ , and Eq. (\ref{2ndFTEq}) is still valid in a harmonic potential as long as one replaces $ F(x) $ with $ -V^{\prime}(x)=-m\omega^{2}x $. We define the second moment in thermal equilibrium as $\langle x^{2}\rangle_{th}=k_{b}T/(m\omega^{2})$. Since $ A $ isn't necessarily positive, $ p $ here is the Fourier pair of $ A $ and Eq. (\ref{2ndFTEq}) can be rewritten as
\begin{equation} \label{eq50}
\begin{split}
\frac{\partial}{\partial t}&G(x,p,t)+pU(x)G(x,p,t)=L_{fp}D_{t}^{1-\alpha,\lambda}G(x,p,t) \\
&+\left[\lambda^{\alpha}D_{t}^{1-\alpha,\lambda}-\lambda\right]\left[G(x,p,t)-e^{-pU(x)t}\delta(x)\right]
\end{split}
\end{equation}
In Laplace space, Eq. (\ref{eq50}) becomes
\begin{equation} \label{eq51}
\begin{split}
&~~~~~sG(x,p,s)-\delta(x)-ipxG(x,p,s)= \\
&K_{\alpha}\left[\frac{\partial^{2}}{\partial x^{2}}+\frac{\partial}{\partial x}\frac{m\omega^{2}x}{k_{b}T}\right](\lambda+s-ipx)^{1-\alpha}G(x,p,s) \\
&+[\lambda^{\alpha}(\lambda+s-ipx)^{1-\alpha}-\lambda]\cdot \left[G(x,p,s)-\frac{\delta(x)}{s-ipx}\right].
\end{split}
\end{equation}
The following relationships will be used in the following calculations,
$$ \langle A^{2}\rangle_{s}=\int_{-\infty}^{+\infty}(-i)^{2}\left(\frac{\partial^{2}}{\partial p^{2}}G(x,p,s)\right)\Big|_{p=0}dx; $$
$$ \langle Ax\rangle_{s}=\int_{-\infty}^{+\infty}(-i)x \left(\frac{\partial}{\partial p}G(x,p,s)\right)\Big|_{p=0}dx; $$
$$ \langle x^{2}\rangle_{s}=\int_{-\infty}^{+\infty}x^{2}G(x,p=0,s)dx; $$
$$ \int_{-\infty}^{+\infty}G(x,p=0,s)dx=\frac{1}{s}. $$
To get $ \langle A^{2}\rangle_{s} $, operating on both sides of Eq. (\ref{eq51}) with $ -\frac{\partial^{2}}{\partial p^{2}} $, letting $ p=0 $, and integrating over all $ x $, we have, in $ s $ space,
\begin{equation} \label{eq52}
\begin{split}
 s\langle A^{2}\rangle_{s}
= & [2-2\lambda^{\alpha}(1-\alpha)(\lambda+s)^{-\alpha}]\langle Ax\rangle_{s} \\
 & +[\lambda^{\alpha}(\lambda+s)^{1-\alpha}-\lambda]\langle A^{2}\rangle_{s} \\
 & -\lambda^{\alpha}\alpha(1-\alpha)(\lambda+s)^{-\alpha-1}\langle x^{2}\rangle_{s},
\end{split}
\end{equation}
where we used the fact that the integral over the Fokker-Plank operator vanishes, since $ x^{n}G(x,p=0,s) $ and $ x^{n}\frac{\partial G(x,p=0,s)}{\partial x} $ are zero for $ |x|\rightarrow \infty $.
To get $ \langle Ax\rangle_{s} $, operating on both sides of Eq. (\ref{eq51}) with $ \frac{\partial}{\partial p} $, substituting $ p=0 $, multiplying by $ -ix $, and integrating over all $ x $, we obtain, in $ s $ space,
$$
\begin{array}{ll}
 s\langle Ax \rangle_{s}
 =& \left[\lambda^{\alpha}(\lambda+s)^{1-\alpha}-\lambda-\frac{K_{\alpha}\frac{m\omega^{2}}{k_{b}T}}{(\lambda+s)^{\alpha-1}}\right]\langle Ax\rangle_{s}\\
&+ \left[1-\frac{\lambda^{\alpha}(1-\alpha)}{(\lambda+s)^{\alpha}}+\frac{K_{\alpha}\frac{m\omega^{2}}{k_{b}T}(1-\alpha)}{(\lambda+s)^{\alpha}}\right]\langle x^{2}\rangle_{s}.
\end{array}
$$
To get $ \langle x^{2}\rangle_{s} $, letting $ p=0 $, multiplying by $ x^{2} $, and integrating over all $ x $, we have, in $ s $ space,
\begin{equation} \label{eq54}
\begin{split}
s\langle x^{2}\rangle_{s}= & \frac{2K_{\alpha}(\lambda+s)^{1-\alpha}}{s}+[\lambda^{\alpha}(\lambda+s)^{1-\alpha}-\lambda]\langle x^{2}\rangle_{s}   \\
& -2K_{\alpha}\frac{m\omega^{2}}{k_{b}T}(\lambda+s)^{1-\alpha}\langle x^{2}\rangle_{s}.
\end{split}
\end{equation}
From Eq. (\ref{eq54}), we get
$$ \langle x^{2}\rangle_{s}=\frac{2K_{\alpha}}{s[(\lambda+s)^{\alpha}+
2K_{\alpha}\frac{m\omega^{2}}{k_{b}T}-\lambda^{\alpha}]}. $$
Then there exists
\begin{equation}
\begin{split}
\langle Ax\rangle_{s}&=\frac{\tau^{\alpha}(\lambda+s)^{\alpha}+(1-\alpha)(1-\tau^{\alpha}\lambda^{\alpha})}
{\tau^{\alpha}(s+\lambda)^{\alpha}+(1-\tau^{\alpha}\lambda^{\alpha})} \\
& \times\frac{2\langle x^{2}\rangle_{th}}{s(s+\lambda)[\tau^{\alpha}(s+\lambda)^{\alpha}+2-\tau^{\alpha}\lambda^{\alpha}]},
\end{split}
\end{equation}
where we defined the relation time $\tau^{\alpha}=k_{b}T/(K_{\alpha}m\omega^{2})=\langle x^{2}\rangle_{th}/K_{\alpha}$. Next, we can obtain $ \langle A^{2}\rangle_{s} $ with the results of $ \langle Ax\rangle_{s} $ and $ \langle x^{2}\rangle_{s} $,
\begin{widetext}\begin{equation}\label{eq55}
\langle A^{2}\rangle_{s} = \frac{2\langle x^{2}\rangle_{th}}{s(s+\lambda)^{2}}\cdot\frac{2\tau^{\alpha}[(s+\lambda)^{\alpha}-\lambda^{\alpha}]^{2}+
[(3\alpha+\alpha^{2})\lambda^{\alpha}\tau^{\alpha}+2-2\alpha][(s+\lambda)^{\alpha}-\lambda^{\alpha}]+
2\alpha^{2}\lambda^{2\alpha}\tau^{\alpha}+\alpha(1-\alpha)\lambda^{\alpha}}{[(s+\lambda)^{\alpha}-\lambda^{\alpha}]
[\tau^{\alpha}((s+\lambda)^{\alpha}-\lambda^{\alpha})+2][\tau^{\alpha}((s+\lambda)^{\alpha}-\lambda^{\alpha})+1]}.
\end{equation}
\end{widetext}
To find the long-times behavior of the fluctuations (\ref{fluctuaions}), we expand Eq. (\ref{eq55}) for small $s$, invert, and divide by $t^{2}$ (for another derivation based on Eq. (\ref{a2stFTEq}), see Appendix C),
\begin{equation}\label{flu2}
\langle(\triangle\overline{x})^{2}\rangle_{t}\simeq \frac{2\alpha\lambda^{\alpha}\tau^{\alpha}+1-\alpha}{\lambda}\cdot\frac{\langle x^{2}\rangle_{th}}{t}.
\end{equation}
Similarly, for short times,
\begin{equation}\label{fluctuaion1}
\langle(\triangle\overline{x})^{2}\rangle_{t}\simeq\frac{4\langle x^{2}\rangle_{th}}{\Gamma(3+\alpha)}\left(\frac{t}{\tau}\right)^{\alpha}.
\end{equation}
Noting that $\langle x^{2}\rangle_{th}/\tau^{\alpha}=K_{\alpha}$, we can rewrite Eq. (\ref{fluctuaion1}) as $\langle(\triangle\overline{x})^{2}\rangle_{t}\simeq\frac{4K_{\alpha}}{\Gamma(3+\alpha)}t^{\alpha}$, which is, as expected, equal to the results in \cite{Carmi:01,Carmi:10}.

\section{Summary}
Since 1949 the distribution of the functionals of the path of a Brownian particle has attracted the interests of scientists. Anomalous diffusion is found to be ubiquitous in nature and is well studied in recent decades. And in more recent years, the fractional Feynman-Kac equations were derived to well describe the functional distribution of the anomalous diffusive paths. The CTRW model, constituting of the random variables of waiting time and jump length, plays central role in characterizing anomalous diffusion. Because of the finite life span of biological particles and the boundedness of physical spaces, sometimes the more reasonable choice for the distributions of the waiting time and jump length is tempered power-law instead of power-law.
In this paper, we use the general Carmi-Barkai formula Eq. (\ref{eq4}) for the functionals of
CTRW paths to obtain the equation recently proposed by Cairoli and Baule \cite{Baule:00}, who also present a nice application of the theory in the context of stochastic calculus. We further derive the tempered fractional Feynman-Kac equations with the power-law jump length distribution, and the tempered power-law jump length distribution.
 And the case involving external potential is also considered. The tempered fractional Feynman-Kac equations describe the functional distribution of the paths of tempered anomalous dynamics.
We present several applications of the tempered fractional Feynman-Kac equations. In the force-free system, we discuss a few functionals of interest, including the occupation time in half-space, the first-passage time and the maximal displacement. For a particle in a harmonic field, we calculate the fluctuations of the time-averaged position under the limit condition.


\section*{Acknowledgments}
This work was supported by the Fundamental Research Funds for the Central Universities under Grant No. lzujbky-2015-77, the National Natural Science Foundation of China under Grant No. 11271173, and the Israel Science Foundation.

\section*{APPENDIX}

\subsection{Derivation of Eq. (\ref{GkpsLatt})}\label{AppendixA}
For a particle to be at $(x,A)$ at time $t$, according to the model, it must have been at $[x,A-\tau U(x)]$ at time $t-\tau$ when the last jump was made. Let $Y(x,A,t)dt$ be the probability of the particle to jump into $(x,A)$ in the time interval $[t,t + dt]$. We obtain \cite{Carmi:01}
\begin{equation}\label{GxAt}
G(x,A,t)=\int_{0}^{t}W(\tau,\lambda)Y[x,A-\tau U(x),t-\tau]d\tau,
\end{equation}
where $W(\tau,\lambda)=1-\int_{0}^{\tau}\psi(\tau',\lambda)d\tau'$ is the probability for not moving in a time interval $(t-\tau,t)$.
Assume that $U(x)\geq 0$ for all $x$ and thus $A\geq 0$. Laplace transforming Eq. (\ref{GxAt}) $A\rightarrow p$, $t\rightarrow s$ by the shift property and convolution theorem, and taking Fourier transform by the well-known Fourier transformation $ \mathcal{F} \{ xf(x);k \}=-i\frac{\partial}{\partial k}\hat{f}(k)$, we have
\begin{equation}\label{Gkps}
G(k,p,s)=\hat{W}\left[s+pU\left(-i\frac{\partial}{\partial k}\right),\lambda\right]\cdot Y(k,p,s).
\end{equation}
The symbol $U(-i\frac{\partial}{\partial k})$ corresponds the original function $U(x)$, but with $-i\frac{\partial}{\partial k}$ as its argument.
To compute $Y$, we notice that to reach at $(x,A)$ at time $t$, the particle must jump from $[x-a,A-\tau U(x-a)]$ or $[x+a,A-\tau U(x+a)]$ after the waiting time $\tau$ with probability 1/2 for each event. Therefore,
\begin{equation}\label{YxAt}
\begin{split}
& Y(x,A,t)=\delta(x)\delta(A)\delta(t) \\
&+\int_{0}^{t}\psi(\tau,\lambda)\frac{1}{2}Y[x+a,A-\tau U(x+a),t-\tau]d\tau \\
&+\int_{0}^{t}\psi(\tau,\lambda)\frac{1}{2}Y[x-a,A-\tau U(x-a),t-\tau]d\tau.
\end{split}
\end{equation}
The term $\delta(x)\delta(A)\delta(t)$ is the initial condition, which means that at $t=0$, $A=0$ and the particle is at $x=0$. Taking Laplace transforms, $A\rightarrow p$, $t\rightarrow s$, and Fourier transform $x\rightarrow k$  of Eq. (\ref{YxAt}), we find
\begin{equation}\label{Ykps}
\begin{split}
& Y(k,p,s)=1 \\
& +\frac{1}{2}e^{-ika}\int_{-\infty}^{\infty}e^{ikx}\psi[s+pU(x),\lambda]\cdot Y(x,p,s)dx \\
& +\frac{1}{2}e^{ika}\int_{-\infty}^{\infty}e^{ikx}\psi[s+pU(x),\lambda]\cdot Y(x,p,s)dx \\
& =1+\cos(ka)\hat{\psi}[s+pU(-i\frac{\partial}{\partial k}),\lambda]\cdot Y(k,p,s).
\end{split}
\end{equation}
Notice that the order of the terms is important, since the $\cos(ka)$ does not commute with $\hat{\psi}[s+pU(-i\frac{\partial}{\partial k}),\lambda]$. This order of operators is natural, because in CTRW we first wait and then make a jump. Rearranging Eq. (\ref{Ykps}), then substituting into Eq. (\ref{Gkps}), we have
$$
\begin{array}{l}
G(k,p,s) \\
=\hat{W}[s+pU(-i\frac{\partial}{\partial k}),\lambda]\cdot \frac{1}{1-\cos(ka)\hat{\psi}[s+pU(-i\frac{\partial}{\partial k}),\lambda]} \\
=\frac{1-\hat{\psi}[s+pU(-i\frac{\partial}{\partial k}),\lambda]}{s+pU(-i\frac{\partial}{\partial k})}
\cdot \frac{1}{1-\cos(ka) \hat{\psi}[s+pU(-i\frac{\partial}{\partial k}),\lambda]},
\end{array}
$$
where we used the fact that $\hat{W}(s,\lambda)=[1-\hat{\psi}(s,\lambda)]/s$.
\begin{widetext}
\subsection{TFRD operator $\nabla_{x}^{\beta,\gamma}$}\label{AppendixD}
Now we give the exact expression of the TFRD operator $ \nabla_{x}^{\beta,\gamma} $.
For $ 0<\beta<1 $,
\begin{equation}\label{TFRD Eq}
\int_{-\infty}^{+\infty}\left(e^{ikx}-1\right)\frac{\beta A_{\beta}}{\Gamma(1-\beta)}e^{-\gamma |x|}|x|^{-\beta-1}dx
 =-A_{\beta}^{\theta}\left(\gamma^{2}+k^{2}\right)^{\beta/2}+2A_{\beta}\gamma^{\beta}.
\end{equation}
Multiplying both sides of Eq. (\ref{TFRD Eq}) by $f(k)$, we get
\begin{equation}
\begin{split}
\left[A_{\beta}^{\theta}\left(\gamma^{2}+k^{2}\right)^{\beta/2}-2A_{\beta}\gamma^{\beta}\right]f(k)
& =2A_{\beta}\cos(\beta\theta)(\gamma^{2}+k^{2})^{\beta/2}f(k)-2A_{\beta}\gamma^{\beta}f(k) \\
& =\int_{-\infty}^{+\infty}\left(f(k)-e^{iky}f(k)\right)\frac{\beta A_{\beta}}{\Gamma(1-\beta)}e^{-\gamma |y|}|y|^{-\beta-1}dy.
\end{split}
\end{equation}
Then using the shift property $ \int e^{ikx}f(x-y)dx=e^{iky}f(k) $ of the Fourier transform leads to
\begin{equation}\label{FRDO}
-2\cos(\frac{\beta\pi}{2})\nabla_{x}^{\beta,\gamma}f(x)-2\gamma^{\beta}f(x)
=\frac{\beta}{\Gamma(1-\beta)}\int_{-\infty}^{+\infty}\frac{f(x)-f(x-y)}{|y|^{\beta+1}}e^{-\gamma|y|}dy.
\end{equation}
From Eq. (\ref{FRDO}), we have
\begin{equation}\label{NABLA}
\begin{split}
\nabla_{x}^{\beta,\gamma}f(x)= & -\frac{1}{2\cos(\frac{\beta\pi}{2})}\left[\frac{\beta}{\Gamma(1-\beta)}
\int_{-\infty}^{+\infty}\frac{f(x)-f(x-y)}{|y|^{\beta+1}}e^{-\gamma|y|}dy+2\gamma^{\beta}f(x)\right] \\
=& -\frac{1}{2\cos(\frac{\beta\pi}{2})}\Big[\frac{\beta}{\Gamma(1-\beta)}\int_{-\infty}^{x}\frac{f(x)-f(y)}{(x-y)^{1+\beta}}e^{-\gamma(x-y)}dy
+\gamma^{\beta}f(x)
\\
&+\frac{\beta}{\Gamma(1-\beta)}\int_{x}^{+\infty}\frac{f(x)-f(y)}{(y-x)^{1+\beta}}e^{-\gamma(y-x)}dy
+\gamma^{\beta}f(x)\Big] \\
 =&-\frac{1}{2\cos(\frac{\beta\pi}{2})}[_{-\infty}\mathbb{D}_{x}^{\beta,\gamma}f(x)+\,_{x}\mathbb{D}_{+\infty}^{\beta,\gamma}f(x)];
\end{split}
\end{equation}
for $ 1<\beta<2 $, repeating the process above leads to
\begin{equation}\label{nabla2}
\begin{split}
\nabla_{x}^{\beta,\gamma}f(x)= & -\frac{1}{2\cos(\frac{\beta\pi}{2})}\left[\frac{\beta(\beta-1)}{\Gamma(2-\beta)}
\int_{-\infty}^{+\infty}\frac{f(x-y)-f(x)+yf^\prime(x)}{|y|^{\beta+1}}e^{-\gamma|y|}dy+2\gamma^{\beta}f(x)\right] \\
= & -\frac{1}{2\cos(\frac{\beta\pi}{2})}\Big[\frac{\beta(\beta-1)}{\Gamma(2-\beta)}
\int_{-\infty}^{x}\frac{f(y)-f(x)+(x-y)f^\prime(x)}{(x-y)^{1+\beta}}e^{-\gamma(x-y)}dy
+\gamma^{\beta}f(x)
\\
& +\frac{\beta(\beta-1)}{\Gamma(2-\beta)}\int_{x}^{+\infty}\frac{f(y)-f(x)+(x-y)f^\prime(x)}{(y-x)^{1+\beta}}e^{-\gamma(y-x)}dy
+\gamma^{\beta}f(x)\Big] \\
=& -\frac{1}{2\cos(\frac{\beta\pi}{2})}[_{-\infty}\mathbb{D}_{x}^{\beta,\gamma}f(x)+\,_{x}\mathbb{D}_{+\infty}^{\beta,\gamma}f(x)],
\end{split}
\end{equation}
where $ _{-\infty}\mathbb{D}_{x}^{\beta,\gamma} $ and $ _{x}\mathbb{D}_{+\infty}^{\beta,\gamma} $ are Riemann-Liouville tempered fractional derivative operators \cite{Sabzikar:00}.
When $\gamma=0$, the operator $\nabla_{x}^{\beta,\gamma}f(x)$ reduces to $\nabla_{x}^{\beta}f(x)=-\frac{1}{2\cos(\frac{\beta\pi}{2})}[_{-\infty}D_{x}^{\beta}f(x)+\,_{x}D_{+\infty}^{\beta}f(x)]$.

\subsection{Another derivation of Eq. (\ref{flu2}) based on Eq. (\ref{a2stFTEq})}
In this section, we derive the fluctuations of the time-averaged position from Eq. (\ref{a2stFTEq}). Let's write the forward equation in $(p,s)$ space for the functional $A=\overline{x}t=\int_{0}^{t}x(\tau)d\tau$ and $x_{0}=0$. Since $A$ is not necessarily positive, replacing $p$ with $-ip$ in Eq. (\ref{a2stFTEq}) leads to
\begin{equation} \label{eq100}
sG(x,p,s)-\delta(x)-ipxG(x,p,s)=
K_{\alpha}\left[\frac{\partial^{2}}{\partial x^{2}}+\frac{\partial}{\partial x}\frac{m\omega^{2}x}{k_{b}T}\right](s-ipx)\frac{G(x,p,s)}{(\lambda+s-ipx)^{\alpha}-\lambda^{\alpha}}.
\end{equation}
Using the equations between Eq. (\ref{eq51}) and (\ref{eq52}) in Sec. \uppercase \expandafter {\romannumeral 4}, we have
$$s\langle A^{2}\rangle_{s}=2\langle Ax\rangle_{s},$$
$$s\langle Ax\rangle_{s}-\langle x^{2}\rangle_{s}=\frac{K_{\alpha}m\omega^{2}}{k_{b}T}\cdot\frac{[(\lambda+s)^{\alpha}-\lambda^{\alpha}]-s\alpha(\lambda+s)^{\alpha-1}}
{[(\lambda+s)^{\alpha}-\lambda^{\alpha}]^{2}}\langle x^{2}\rangle_{s}-\frac{K_{\alpha}m\omega^{2}}{k_{b}T}\cdot\frac{s}{[(\lambda+s)^{\alpha}-\lambda^{\alpha}]}\langle Ax\rangle_{s},$$
$$s\langle x^{2}\rangle_{s}=\frac{2K_{\alpha}}{(\lambda+s)^{\alpha}-\lambda^{\alpha}}-\frac{2K_{\alpha}m\omega^{2}}{k_{b}T}\cdot
\frac{s}{(\lambda+s)^{\alpha}-\lambda^{\alpha}}\langle x^{2}\rangle_{s}.$$
From the last three equations, we get
\begin{equation}\label{eq200}
\langle A^{2}\rangle_{s}=\frac{4\langle x^{2}\rangle_{th}}{s^{3}}\cdot\frac{\tau^{\alpha}[(s+\lambda)^{\alpha}-\lambda^{\alpha}]^{2}+
(s+\lambda)^{\alpha}-\lambda^{\alpha}-s\alpha(\lambda+s)^{\alpha-1}}
{[(s+\lambda)^{\alpha}-\lambda^{\alpha}]
[\tau^{\alpha}((s+\lambda)^{\alpha}-\lambda^{\alpha})+2][\tau^{\alpha}((s+\lambda)^{\alpha}-\lambda^{\alpha})+1]}.
\end{equation}
To find the long-times behavior of the fluctuations (\ref{fluctuaions}), we expand Eq. (\ref{eq200}) for small $s$, invert, and divide by $t^{2}$,
$$\langle(\triangle\overline{x})^{2}\rangle_{t}\simeq \frac{2\alpha\lambda^{\alpha}\tau^{\alpha}+1-\alpha}{\lambda}\cdot\frac{\langle x^{2}\rangle_{th}}{t},$$
which is the same as Eq. (\ref{flu2}).

\end{widetext}

\end{document}